\documentclass[apj]{emulateapj}

\usepackage{natbib}
\usepackage{color}
\usepackage{graphicx}
\usepackage{epstopdf}

\newcommand{\Mjup}{\ensuremath{\mathrm{M}_{\mathrm{Jup}}\!}}
\newcommand{\Msun}{\ensuremath{\mathrm{M}_{\odot}\!}}

\newcommand{\ipfilt}{\ensuremath{i^{\prime}}}
\newcommand{\zpfilt}{\ensuremath{z^{\prime}}}
\newcommand{\Kfilt}{\ensuremath{K}}
\newcommand{\Kpfilt}{\ensuremath{K^{\prime}}}
\newcommand{\Ksfilt}{\ensuremath{K_s}}
\newcommand{\Kcfilt}{\ensuremath{K_c}}
\newcommand{\BrGfilt}{\ensuremath{BrG}}
\newcommand{\Hfilt}{\ensuremath{H}}

\newcommand{\Jfilt}{\ensuremath{J}}
\newcommand{\Jcfilt}{\ensuremath{J_c}}

\renewcommand{\deg}{\ensuremath{^{\circ}}}
\renewcommand{\arcsec}{\ensuremath{^{\prime\prime}}}

\newcommand{\unit}[1]{\ensuremath{\,\textrm{#1}}}

\slugcomment{Accepted for publication in ApJ}

\shorttitle{Friends of hot Jupiters IV}
\shortauthors{Ngo et al.}

\begin{document}


\title{Friends of Hot Jupiters. IV. Stellar companions beyond 50 AU might facilitate\\ giant planet formation, but most are unlikely to cause Kozai-Lidov migration}
 

\author{Henry Ngo\altaffilmark{1}, Heather A. Knutson\altaffilmark{1}, Sasha Hinkley\altaffilmark{2}, Marta Bryan\altaffilmark{3}, Justin R. Crepp\altaffilmark{4}, Konstantin Batygin\altaffilmark{1}, \\Ian Crossfield\altaffilmark{5}, Brad Hansen\altaffilmark{6}, Andrew W. Howard\altaffilmark{7}, John A. Johnson\altaffilmark{8}, Dimitri Mawet\altaffilmark{3,9}, \\Timothy D. Morton\altaffilmark{10}, Philip S. Muirhead\altaffilmark{11}, and Ji Wang\altaffilmark{3}}
\email{hngo@caltech.edu}

\altaffiltext{1}{Division of Geological and Planetary Sciences, California Institute of Technology, Pasadena, CA, USA}
\altaffiltext{2}{Department of Physics and Astronomy, University of Exeter, Exeter, UK}
\altaffiltext{3}{Department of Astronomy, California Institute of Technology, Pasadena, CA, USA}
\altaffiltext{4}{Department of Physics, University of Notre Dame, Notre Dame, IN, USA}
\altaffiltext{5}{Lunar and Planetary Laboratory, University of Arizona, Tucson, AZ, USA}
\altaffiltext{6}{Department of Physics and Astronomy, University of California Los Angeles, Los Angeles, CA, USA}
\altaffiltext{7}{Institute for Astronomy, University of Hawaii at Manoa, Honolulu, HI, USA}
\altaffiltext{8}{Harvard-Smithsonian Center for Astrophysics, Cambridge, MA, USA}
\altaffiltext{9}{Jet Propulsion Laboratory, California Institute of Technology, Pasadena, CA, USA}
\altaffiltext{10}{Department of Astrophysical Sciences, Princeton University, Princeton, NJ, USA}
\altaffiltext{11}{Department of Astronomy, Boston University, Boston, MA, USA}


\begin{abstract}
Stellar companions can influence the formation and evolution of planetary systems, but there are currently few observational constraints on the properties of planet-hosting binary star systems. We search for stellar companions around 77 transiting hot Jupiter systems to explore the statistical properties of this population of companions as compared to field stars of similar spectral type. After correcting for survey incompleteness, we find that $47\%\,\pm\,7\%$ of hot Jupiter systems have stellar companions with semi-major axes between 50\unit{AU}-2000\unit{AU}. This is 2.9 times larger than the field star companion fraction in this separation range, with a significance of $4.4\sigma$. In the 1\unit{AU}-50\unit{AU} range, only $3.9^{+4.5}_{-2.0}\%$ of hot Jupiters host stellar companions compared to the field star value of $16.4\%\,\pm\,0.7\%$, which is a $2.7\sigma$ difference. We find that the distribution of mass ratios for stellar companions to hot Jupiter systems peaks at small values and therefore differs from that of field star binaries which tend to be uniformly distributed across all mass ratios. We conclude that either wide separation stellar binaries are more favorable sites for gas giant planet formation at all separations, or that the presence of stellar companions preferentially causes the inward migration of gas giant planets that formed farther out in the disk via dynamical processes such as Kozai-Lidov oscillations. We determine that less than 20\% of hot Jupiters have stellar companions capable of inducing Kozai-Lidov oscillations assuming initial semi-major axes between 1-5\unit{AU}, implying that the enhanced companion occurrence is likely correlated with  environments where gas giants can form efficiently.
\end{abstract}




\section{Introduction}
Almost half of all FGK stars are in multiple systems~\citep{Raghavan2010}. Therefore, it is important to understand the role that stellar companions play in the formation and evolution of planetary systems. In addition, ongoing transit surveys have demonstrated that a majority of apparently single stars host planets, and have provided unprecedented new opportunities to compare the properties of planets located in binary star systems to those of single stars~\citep{Winn2015}.

The recent proliferation of high contrast imaging of planet hosting stars is closely linked with the {\it Kepler} mission, as this survey was the first to produce large numbers of transiting planet candidates for which radial velocity confirmation was impractical. For these systems, high contrast imaging is required in order to eliminate astrophysical false positives and to correct for dilution of transit light curves. Prior to {\it Kepler}, the first reports of stellar companions came from serendipitous discoveries from newly obtained high contrast images or archival images reported along with the planet discovery~\citep[e.g.][]{CollierCameron2007}. Then, ``Lucky imaging'' techniques~\citep[e.g.][]{Daemgen2009} used adaptive optics (AO) to perform systematic surveys with small sample sizes and modest sensitivity. More recently, there have been a series of larger AO surveys targeting {\it Kepler} planet candidate host stars using state-of-the-art methods and large telescopes to perform diffraction-limited imaging, allowing for better survey sensitivity, especially at short wavelengths~\citep[e.g.][]{Adams2012,Dressing2014,Wang2014}. A full review of these campaigns can be found in~\citet{Ngo2015}.

In this work, we continue the search for stellar companions in systems with hot Jupiters transiting FGK stars in order to explore the potential role of these companions in planet formation and migration. The ``Friends of Hot Jupiters'' (FOHJ) campaign~\citep{Knutson2014,Ngo2015,Piskorz2015}, has searched for planetary and stellar companions to a sample of 50 hot Jupiter hosts via radial velocity monitoring~\citep{Knutson2014}, infrared spectral model comparison~\citep{Piskorz2015} and direct imaging~\citep{Ngo2015}. This original survey sample contained two subpopulations: stars that host planets with some dynamical signature of multi-body interactions, such as a measured offset between the orientation of the planet's orbit and the host star's spin axis or a non-zero orbital eccentricity, and stars that host planets on well-aligned orbits and with orbital eccentricities consistent with zero to three sigma. Our direct imaging survey was the first to apply a statistical approach to estimate the fraction of hot Jupiter host stars with gravitationally bound stellar companions including a correction for survey sensitivity. We found a stellar companion rate of $48\% \pm 9\%$ in the 50-2000\unit{AU} region, showing moderately significant ($2.8\sigma$) evidence for a larger companion fraction around solar-type hot Jupiter hosts than solar-type field stars. Our survey was also the first to systematically examine a sample of planets with spin-orbit measurements, allowing us to compare misaligned and well-aligned systems. We found no evidence for a correlation between the presence of an outer stellar or planetary companion in these systems and the orbital properties of the inner transiting hot Jupiter. 

More recently, there have been four large direct imaging surveys for companions to transiting gas giant planet hosts~\citep{Woellert2015a,Woellert2015b,Wang2015b,Evans2016}. \citet{Woellert2015a} applied stellar density arguments to estimate that 12 out of their 49 targets have bound companions while \citet{Woellert2015b} report candidate companions around 33 out of 74 systems. Although these studies do not confirm common proper motion or report a survey sensitivity corrected companion rate, their raw companion fractions are consistent with ours. \citet{Wang2015b} and \citet{Evans2016} did check for common proper motion and correct for survey sensitivity. \citet{Wang2015b} report a stellar multiplicity rate for {\it Kepler} hot Jupiter hosts to be $51\%\,\pm\,13\%$ and \citet{Evans2016} found a companion rate of $38^{+17}_{-13}\%$. Both of these results are in good agreement with our previously published value.

Although the higher binary fraction of hot Jupiter host stars suggests these stellar companions play a role in the creation of hot Jupiters, it is unclear exactly what this role might be. In one class of scenarios, the presence of a stellar companion might cause gas giant planets formed at larger separations to migrate inward via secular interactions such as the Kozai-Lidov effect~\citep[e.g.][]{Fabrycky2007,Naoz2012,Naoz2013,Storch2014,Dawson2015,Petrovich2015a,Anderson2016,Munoz2016}. If stellar Kozai is the dominant migration mechanism, it should result in a population of hot Jupiters with a broad distribution of orbital inclinations that is closely correlated with the presence of companions. However, our earlier direct imaging survey finds no correlation between the orbital properties of the transiting planet and stellar multiplicity, suggesting that Kozai-Lidov migration is probably not the dominant channel for the generation of hot Jupiter spin-orbit misalignment. Instead, our results signal broad agreement with the primordial excitation of stellar obliquities~\citep[e.g.,][]{Spalding2014,Spalding2015,Lai2014,Fielding2015}.

In an alternative scenario, we consider the possibility that stellar binaries are more favorable locations for the formation of gas giant planets. Some previous studies suggested that stellar companions might suppress gas giant planet formation by exciting planetesimal velocity dispersions~\citep{Mayer2005}, truncating the disk~\citep{Pichardo2005,Kraus2012,Cheetham2015}, or ejecting newly formed planets~\citep{Kaib2013,Zuckerman2014}. Other theoretical studies, however, have shown that disk self-gravity successfully shields planet-formation environments from companion-driven secular excitation of embedded orbits~\citep{Batygin2011,Rafikov2013b}. The observed enhanced binary rate for hot Jupiter host stars suggests that planet formation is indeed unhindered in these systems.

In this study we increase the sample size of our direct imaging survey from 50 transiting hot Jupiter systems to 77 systems in order to take a closer look at the properties of the observed population of stellar companions and to place improved constraints on the possible effects of these companions on hot Jupiter formation. We obtain a more precise measurement of hot Jupiter stellar multiplicity and characterize the mass ratio as well as semi-major axis distributions of the observed population of companions as compared to those of solar-type field stars. Finally, while our previous work shows that hot Jupiter migration via Kozai-Lidov oscillations is unlikely, this work uses the larger sample size to place quantitative upper limits on this migration mechanism.

This paper is structured as follows. Section~\ref{sec:obs} describes our observations. In Section~\ref{sec:analysis} we characterize companion properties and determine our contrast limits. Section~\ref{sec:companions} describes each of the individual multistellar systems detected in our new observations. Section~\ref{sec:surveyresults} reports our survey results, companion rates, and trends in the properties of the observed population of stellar companions. Section~\ref{sec:discuss} discusses the implications of our results for hot Jupiter planet formation and constrains the fraction of systems affected by Kozai-Lidov. Section~\ref{sec:summary} presents a summary of this work.

\section{Sample selection and observations}
\label{sec:obs}
Our total sample consists of 82 systems known to host transiting gas giant planets. We divide our sample into two populations. The first population, containing 77 stars, is our ``survey sample'', which is the only population we use in all of the estimates of hot Jupiter companion fraction and other constraints presented in this work. The first 50 targets in this sample are the same set of stars used in the first three FOHJ papers. For more information on the selection of these targets, see~\citet{Knutson2014}. The remaining 27 targets are new systems with transiting gas giant planets with masses between 0.27\unit{\Mjup} and 4.06\unit{\Mjup} and separations between 0.014\unit{AU} and 0.061\unit{AU}.  They were selected without regard to whether or not the stars had directly imaged stellar companions reported by other imaging surveys.  We also relax our previous preference for systems with published spin-orbit alignment measurements, as our initial survey results found no evidence for any correlation between this parameter and the presence of a stellar companion.

The second population is a set of five targets (HAT-P-54, WASP-36, WASP-58, WASP-76, WASP-103) that we decided to observe only after their stellar companions were reported in the published literature~\citep{Woellert2015a,Woellert2015b}. Therefore, they do not form a part of our survey population and we exclude them from our statistical analysis discussed in Section~\ref{sec:surveyresults}. We characterize the companions around these non-survey targets following the same procedure as the survey targets, to be described in Section~\ref{sec:analysis}, and report on these systems individually in Section~\ref{sec:companions}. Although these targets cannot be fairly considered in our determination of the hot Jupiter companion rate, we are still able to confirm the existence of the companions around non-survey targets from previous studies and provide new or updated companion properties.

We obtained $K$ band AO observations using the NIRC2 instrument (instrument PI: Keith Matthews) on Keck II between February 2012 and January 2016. These new observations are summarized in Table~\ref{tab:obs}. We follow the same procedure described in \citet{Ngo2015}. We operated in the natural guide star mode using the narrow camera setting, which yields a plate scale of 10$\,$mas$\,$pixel$^{-1}$. Each survey target had at least one series of $K$ band observations with at least 105 seconds of on-sky integration time. As in our previous survey, this strategy allows us to reach contrasts of $\Delta K$ of 8 magnitudes at $1''$ of separation. For targets where a companion was detected, we also take observations in $J$ and/or $H$ bands in order to obtain a measurement of the companion's color. We also test for common proper motion using additional epochs of $K$ band imaging obtained 1-3 years after the initial detection. These followup photometric and astrometric observations may have shorter integration time.

We use dome flats and dark frames to calibrate our images and to identify hot pixels and dead pixels using the criteria described in \cite{Ngo2015}. We utilize these individual calibrated frames for our photometric and astrometric analysis, while we perform our sensitivity calculations on the median stack of these individual frames.

\begin{figure*}
\epsscale{1.0}
\plotone{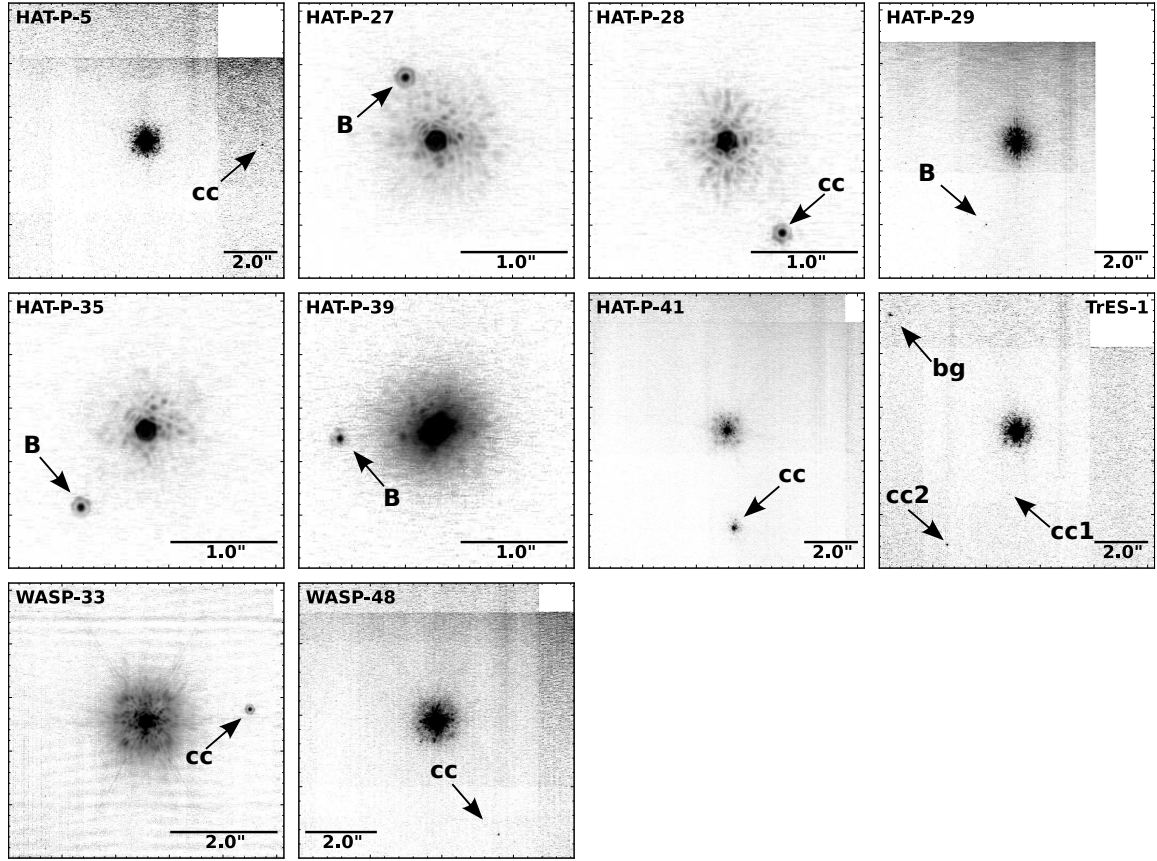}
\caption{Median-stacked K band image for each detected candidate mutli-stellar system presented in this work, from our survey targets. Each image is oriented such that North points up and East to the left. 
\label{fig:comps_survey}}
\end{figure*}

\begin{figure*}
\epsscale{0.8}
\plotone{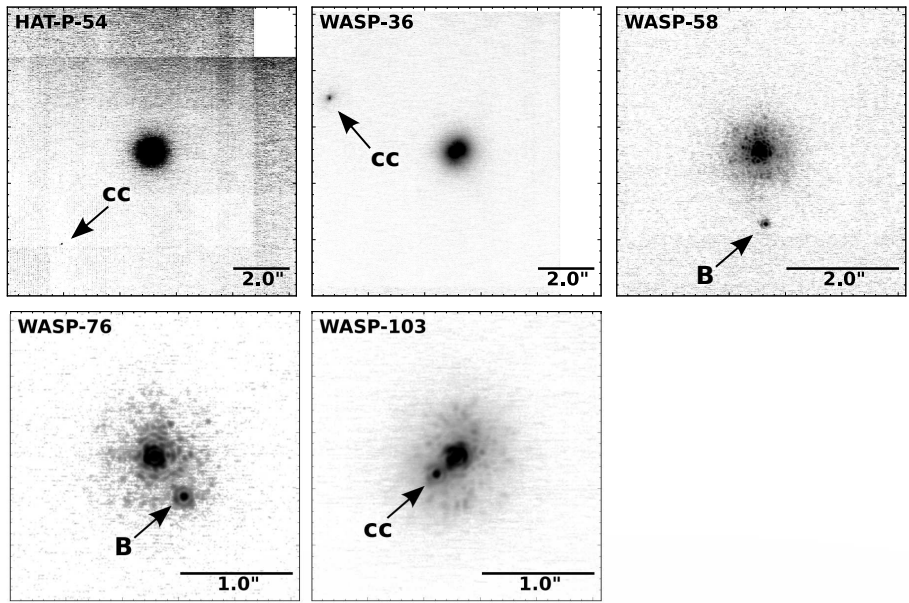}
\caption{Median-stacked K band image for each detected candidate mutli-stellar system presented in this work, from our non-survey targets. Each image is oriented such that North points up and East to the left. 
\label{fig:comps_nonsurvey}}
\end{figure*}

\section{Analysis of Companion Properties}
\label{sec:analysis}
\subsection{PSF fitting}
\label{sec:psffit}
We identify candidate companions around 15 of our target stars (see Figures~\ref{fig:comps_survey} and \ref{fig:comps_nonsurvey}). We summarize the stellar parameters for all observed stars in Table~\ref{tab:stellar_params}.

We measure the flux ratio and on-sky separation for each detected multi-stellar system by fitting each image with a multiple-source point spread function (PSF) modelled as a combination of a Moffat and Gaussian functions. For the functional form and description of the parameters, see~\citet{Ngo2015}. We use a maximum likelihood estimation routine to find the best fit parameters and create an analytic form for our PSF model using these parameters. Integrating this PSF model over a circular aperture for each star yields the flux ratio. The difference in the stellar position parameters determines the separation as projected onto the NIRC2 array. To get the true on-sky separation and position angle between the stars, we use the known NIRC2 astrometric corrections~\citep{Yelda2010,Service2016submitted} to account for the NIRC2 distortion and rotation\footnote{The \citet{Yelda2010} was used for NIRC2 data taken prior to 2015 April 13. Realignment of the Keck2 AO bench caused a change in the NIRC2 distortion solution, so we use the new solution presented by \citet{Service2016submitted} for data taken after this date.}. These astrometric corrections include uncertainties on the distortion, plate scale and orientation of the NIRC2 array and we include all of these uncertainties in our reported errorbars for our measured separation and position angle.

For each individual calibrated frame, we compute the flux ratio and separations as outlined above. We then report the best estimate for each of these values as the mean value from all of the frames. We estimate our measurement error as the standard error on the mean. 

We report the best fitting flux ratio between primary and companion stars as a magnitude difference in each survey bandpass in Table~\ref{tab:comp_phot}. We also use apparent magnitudes of the primary star from the 2MASS catalog~\citep{Skrutskie2006} to compute the apparent magnitudes of the companion stars in all bands. Tables~\ref{tab:phot_colors} and~\ref{tab:comp_astr} report all computed photometry and K-band astrometry, respectively, of our detected companion stars. 

\subsection{Common proper motion confirmation}
We are interested in determining whether or not our detected companion stars are gravitationally bound to the primary star. For our candidate mutli-stellar systems, we followed up with K-band images to verify that the companion star shares common proper motion with the primary star. Following the procedure described in~\citet{Ngo2015}, we calculate the evolution of the companion's separation and position angle if it were a background object and compare it to the actual measured separation and position angle at each observation date in Figures~\ref{fig:astr_plot1} and \ref{fig:astr_plot2}. When our candidate companions have been imaged in other surveys and these other surveys report a separation and position angle with uncertainties, we also include these previous measurements. Table~\ref{tab:comp_astr} lists all the astrometric measurements used in our analysis. 

\begin{figure*}
\epsscale{1.18}
\plotone{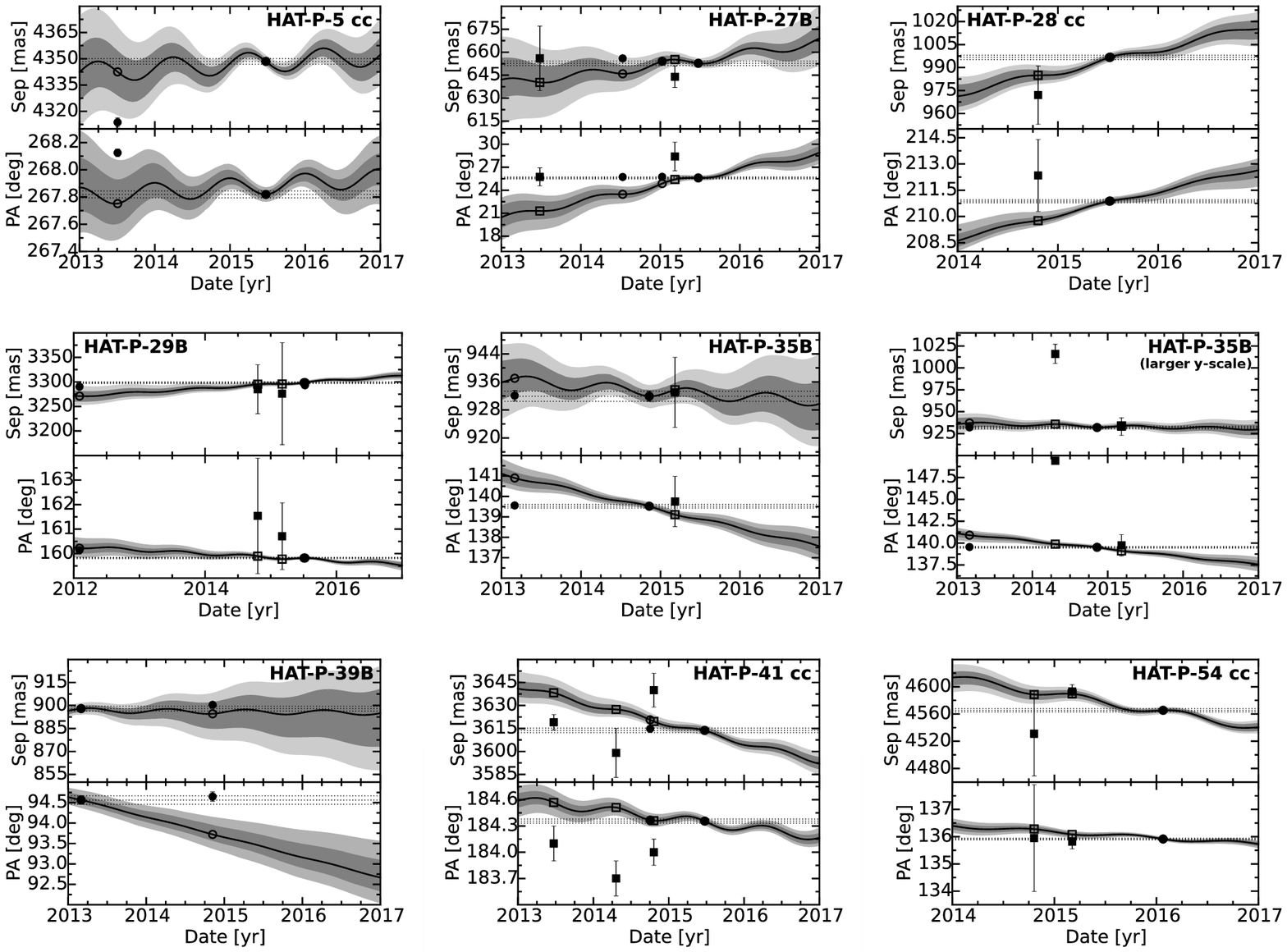}
\caption{Common proper motion confirmation for each candidate companion. The top and bottom panels show the separation and position angle of a candidate companion relative to the primary star. The background track (solid line) starts at the observation with the smallest uncertainty in separation and position angle. The shaded region indicate the 68\% and 95\% confidence regions. We use uncertainties in our separation and position angle measurement as well as the uncertainties in the primary star's celestial coordinates, proper motion, and parallax in our Monte Carlo routine to determine these confidence regions. The filled symbols show measured positions of companions (listed in Table~\ref{tab:comp_astr} and open symbols show the expected position if the candidate object were a very distant background object. Circles are measurements from our campaign while squares are measurements from other studies. When the solid symbols and open symbols differ and when the measurement values are consistent with each other at all observation epochs, then we can conclude our detected object is a physically bound companion. Objects labeled as ``B'' have common proper motion, as ``cc'' are candidate companions and as ``bg'' are background objects. See Section~\ref{sec:companions}. Continued in Figure~\ref{fig:astr_plot2}.
\label{fig:astr_plot1}}
\end{figure*}

\begin{figure*}
\epsscale{1.18}
\plotone{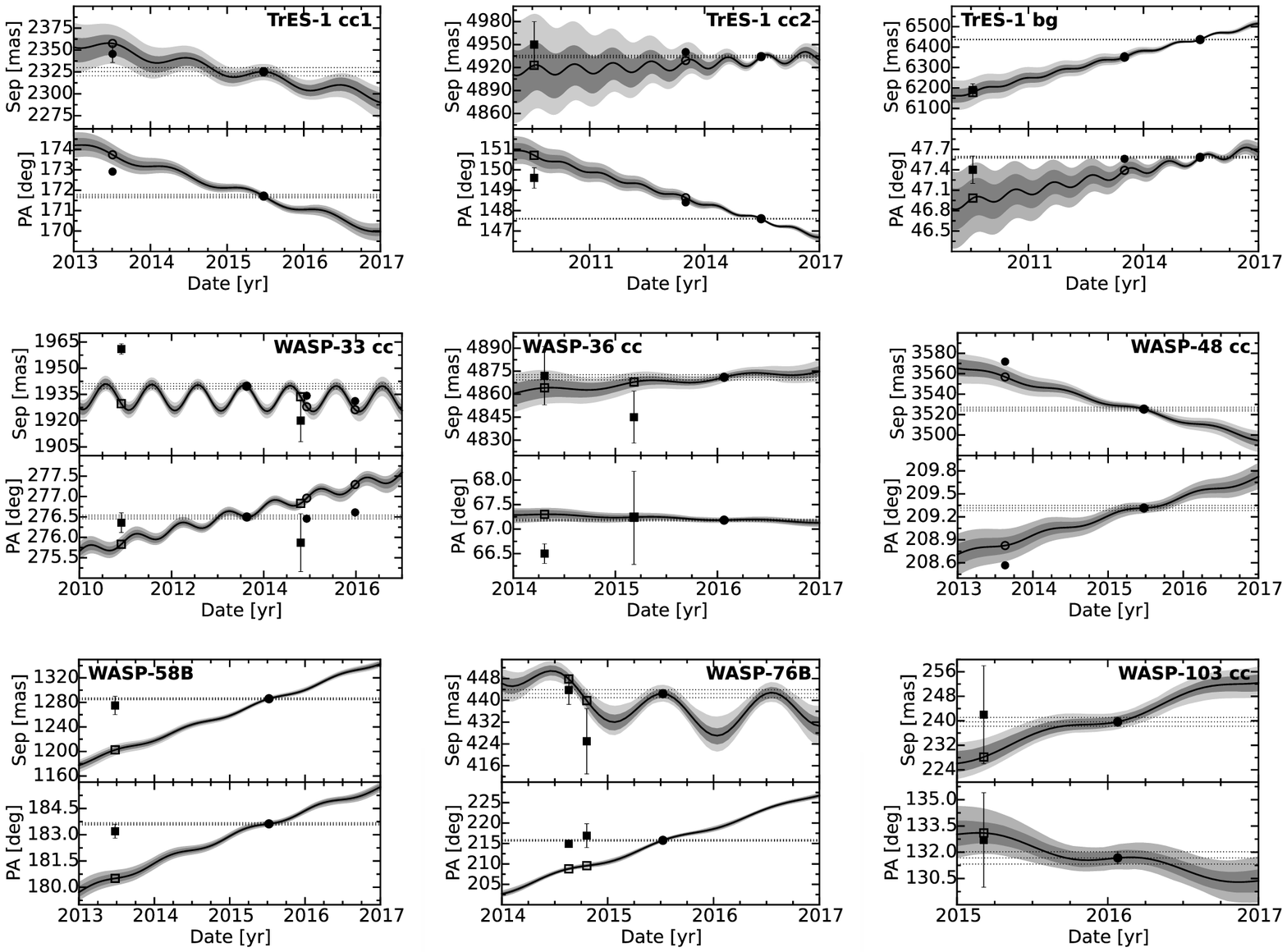}
\caption{Continued from Figure~\ref{fig:astr_plot1}.
\label{fig:astr_plot2}}
\end{figure*}

\subsection{Masses and separation}
For each confirmed multi-stellar system, we compute the companion star's physical parameters using the method described in~\citet{Ngo2015}. In brief, we model the primary and companion star fluxes by integrating the PHOENIX synthetic spectra~\citep{Husser2013} over the observed bandpass. We use the set of models corresponding to solar metallicities and composition ([Fe/H]$=0$, [$\alpha$/H]$=0$). For the primary star, we use previously published measurement of stellar mass, radius, effective temperature and distance as listed in Table~\ref{tab:stellar_params}. For the companion star, we use the same distance measurement and calculate the companion star effective temperature that would result in a companion star flux that matches the observed flux ratio. We use the zero-age main sequence models from \citet{Baraffe1998} to determine the companion star's mass and radius from the effective temperature. Our error budget includes all relevant measurement uncertainties but does not include any model dependent uncertainties from the PHOENIX spectra or the zero-age main sequence model. We calculate effective temperatures for each candidate companion based on the measured flux ratios in the $J$, $H$, and $K$ bands, and ask whether the brightness ratios across all three bands are consistent with the same stellar effective temperature. We report these individual effective temperatures values as well as the average across all three bands in Table~\ref{tab:interp_sec}.

The projected spatial separations are computed using our measured projected on-sky separations and the stellar distance. Because the majority of our stars do not have measured parallaxes, we use a spectroscopic distance estimated derived from the spectral type and the star's apparent magnitude.

\subsection{Contrast curves}
\label{sec:cc}
\begin{figure}
\epsscale{1.2}
\plotone{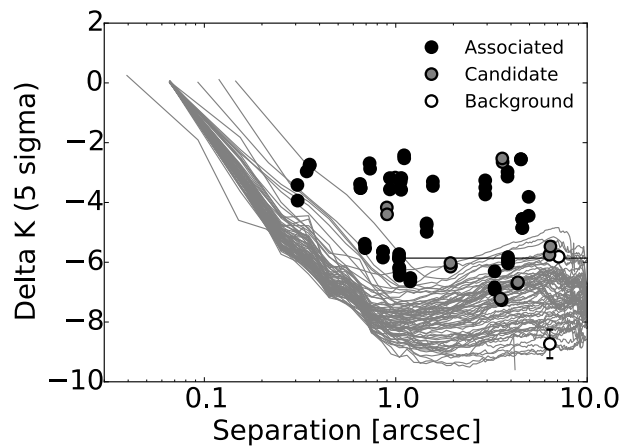}
\caption{$5\sigma$ K band contrast curve computed from stacked images for all observed targets. The curve with the best contrast for each target is shown. For these curves, all companion stars are masked out. Detections of bound companions, candidate companions, and background objects as overplotted as filled black, grey, and open circles, respectively.
\label{fig:contrast}}
\end{figure}

We calculate contrast curves for all targets imaged, regardless of whether or not a companion was detected. Our algorithm is described in \citet{Ngo2015}. Figure~\ref{fig:contrast} shows the K-band $5\sigma$ contrast limit, in magnitudes, for all targets discussed in this paper. We are able to reach a $5\sigma$ contrast of $\Delta K=8$ for most of the targets surveyed. When considering our survey's sensitivity for each target we use its individual contrast curve as discussed in Section~\ref{sec:incomplete}, below.

\section{Notes on detected companions}
\label{sec:companions}
We find 17 candidate stellar companions around 15 of the systems observed, of which 3 are reported for the first time in this paper. In this section, we discuss each system individually and categorize them according to whether or not the companion is bound or not as confirmed by common proper motion measurements. For targets where our astrometric measurements are inconclusive we consider whether or not the companion has colors consistent with the expected spectral type for a bound companion. Our analysis confirms 6 companions as gravitationally bound, identifies 10 candidate companions with inconclusive astrometric measurements and colors consistent with those of a bound companion, and finds 1 candidate companion to be a background object. For each candidate companion, we report the differential magnitude $\Delta K$, separation $\rho$ and position angle (PA) from our first detection epoch for comparison with detections from other studies. We also discuss any observations previously reported by other studies. 

\subsection{Bound companions}

\subsubsection{HAT-P-27 (WASP-40)}
We find a companion with $\Delta \Ksfilt = 3.52 \pm 0.05, \rho = 0\arcsec.656 \pm 0\arcsec.002$ and PA $=25\deg.5 \pm 0\deg.1$. \citet{Woellert2015b} also found a candidate stellar companion at the same separation, however they note that the companion was too dim for them to reliably measure its flux. Our three astrometric measurements show that this companion is physically bound, an argument that is strengthened by the inclusion of the single epoch of astrometry by \citet{Woellert2015b}. \citet{Evans2016} also imaged this system but reported that this companion is below their survey sensitivity.

\subsubsection{HAT-P-29}
We find a companion with $\Delta \Kpfilt = 6.9 \pm 0.2, \rho = 3\arcsec.290 \pm 0\arcsec.002$ and PA $=159\deg.89 \pm 0\deg.03$. Due to our dithering pattern, this faint companion only appeared in a subset of our dithered images. As a result, we failed to identify it in our original 2012 images. After~\citet{Woellert2015a} pointed this out, we revisited our old observations and found that the companion was indeed present in a subset of the frames. Inspection of the contrast curve for this system from \citet{Ngo2015} confirms that the companion fell below our formal $5\sigma$ detection limit, and is therefore consistent with the non-detection reported in \citet{Ngo2015}. We obtained new images of the system in 2015, in which we planned our dither pattern to make sure that the companion remained in the frame in all images. Although the \citet{Woellert2015a} astrometric uncertainties were too large to verify common proper motion, our measurements from 2012 and 2015 show the candidate is consistent with a bound stellar companion.

\subsubsection{HAT-P-35}
We find a companion with $\Delta \Ksfilt = 3.19 \pm 0.06, \rho = 0\arcsec.932 \pm 0\arcsec.002$ and PA $=139\deg.31 \pm 0\deg.09$. \citet{Woellert2015b} also found a companion at the same position but could not confirm common proper motion with only one epoch. Our measurements in 2013 and 2014 confirm this candidate as bound stellar companion. \citet{Evans2016} report a companion in their 2014 images with a similar brightness difference but with separation $\rho  = 1\arcsec.016 \pm 0\arcsec.011$ and PA $=149\deg.4 \pm 0\deg.2$. This measurement is discrepant at the $10\sigma$ level to both of our measurements and at $7\sigma$ to the \citet{Woellert2015b} measurement.

\subsubsection{HAT-P-39}
We find a companion with $\Delta \Ksfilt = 4.2 \pm 0.1, \rho = 0\arcsec.898 \pm 0\arcsec.002$ and PA $=94\deg.3 \pm 0\deg.1$. Our observations in early 2013 and late 2014 show that this candidate companion has the same proper motion as the primary star. The color of this candidate companion is also consistent with a late-type main sequence star. We therefore consider this candidate to be a bound stellar companion.

This system was also imaged by \citet{Woellert2015a} but they did not report a companion. Their detection limit at 1\arcsec\ was $\Delta \zpfilt =4.9$. Our temperature estimate indicates the companion is an early M dwarf, therefore, this candidate may have been below the detection limit of these observations.

\subsubsection{WASP-58}
We find a companion with $\Delta \BrGfilt = 4.4 \pm 0.1, \rho = 1\arcsec.281 \pm 0\arcsec.002$ and PA $=183\deg.37 \pm 0\deg.07$. This companion was originally reported in \citet{Woellert2015a}, and when we combine our single epoch of imaging with the single epoch from their paper we find clear evidence that this candidate is a gravitationally bound companion.

\subsubsection{WASP-76}
We find a companion with $\Delta \BrGfilt = 2.7 \pm 0.1, \rho = 0\arcsec.441 \pm 0\arcsec.002$ and PA $=215\deg.6 \pm 0\deg.2$. This companion was first discovered by \citet{Woellert2015b} and also followed up by \citet{Ginski2016}. When combined with the single-epoch astrometry from these two papers our new epoch of astrometry indicates that this companion is gravitationally bound.

\subsection{Candidate companions}

\subsubsection{HAT-P-5}
We find a companion with $\Delta \Ksfilt = 6.7 \pm 0.2, \rho = 4\arcsec.314 \pm 0\arcsec.003$ and PA $= 267\deg.83 \pm 0\deg.03$. Our astrometric analysis is not well matched by models for either a bound companion or a background object. Because the color of this candidate companion is consistent with a late-type main sequence star, we tentatively consider HAT-P-5 to be a candidate multi-stellar system for the our companion fraction analysis.

This system was also imaged by \citet{Daemgen2009} and \citet{Faedi2013}. \citet{Daemgen2009} did not find this companion, but they restricted their binary search to companions within 2\arcsec. \citet{Faedi2013} noted a potential companion around HAT-P-5 with a separation of $4\arcsec.25$ and position angle of $266\deg$, but classified it as a non-detection because the companion's brightness was below their $4\sigma$ detection limit. We do not use their astrometric point in our analysis because there is no uncertainty reported on their separation. 

\subsubsection{HAT-P-28}
We find a companion with $\Delta \Ksfilt = 3.17 \pm 0.04, \rho = 0\arcsec.994 \pm 0\arcsec.002$ and PA $=210\deg.7 \pm 0\deg.1$. \citet{Woellert2015b} previously reported a candidate stellar companion at a position consistent with our measurement. We include this previous astrometric measurement but it is not precise enough to allow us to distinguish between comoving and bound tracks. Since both our study and \citet{Woellert2015b} find the color of the candidate companion to be consistent with a late type main sequence star, we consider this to be a candidate multi-stellar system in our analysis.

\subsubsection{HAT-P-41}
We find a companion with $\Delta \Ksfilt = 2.65 \pm 0.08, \rho = 3\arcsec.615 \pm 0\arcsec.002$ and PA $=184\deg.10 \pm 0\deg.03$. \citet{Hartman2012} reported a candidate companion along with the discovery of HAT-P-41b at a similar separation, however they do not report a position angle. \citet{Woellert2015a}, \citet{Woellert2015b}, and \citet{Evans2016} all report finding a companion at a similar position. When all observations are taken in account, the astrometric measurements are not well-matched by models for either a bound companion or a background object. Our companion color and effective temperature as well the previous studies' color measurements indicate this companion is consistent with a late type main sequence star at the same distance as the primary star. So, we consider HAT-P-41 to be a candidate multi-stellar system.

\subsubsection{HAT-P-54}
We find a companion with $\Delta \Ksfilt = 6.5 \pm 0.2, \rho = 4\arcsec.557 \pm 0\arcsec.003$ and PA $=135\deg.54 \pm 0\deg.03$. Because the central star has a spectral type of K7, the measured flux ratio predicts a companion temperature below $2300\,\unit{K}$, the lower limit on the PHOENIX models. Therefore, we used a blackbody to model the spectral energy distribution of both the central star and companion. This candidate companion was originally reported in \citet{Woellert2015b}. Their 2014 measurements are consistent with both our 2016 measurement and the background track. Their 2015 measurement has a separation measurement that differs from ours by $3\sigma$ but a consistent position angle. Our measured $\Delta \Ksfilt$ magnitudes corresponds to an effective temperature of $1941\unit{K}\pm75\unit{K}$ for this candidate companion, indicating that it may be a brown dwarf. The $\Delta i^\prime$ and $\Delta z^\prime$ measurements from \citet{Woellert2015b} are also consistent with a brown dwarf candidate.

\subsubsection{TrES-1}
We find three objects around TrES-1. The object closest to the primary has $\rho = 2\arcsec.340 \pm 0\arcsec.01$ and PA $=172\deg.9 \pm 0\deg.1$. This object is too faint for us to get a reliable flux measurement. Our images show a range of differential magnitudes between $\Delta \Ksfilt$ from 7.5 to 9.0. \citet{Adams2013} imaged this system in 2011 and reported a companion with $\Delta \Ksfilt = 7.7, \rho = 2\arcsec.31$ and PA $=174\deg$. Although they do not report any uncertainties, these photometric and astrometric values are consistent with our detection. They also do not detect any additional objects. Our two epochs are consistent with neither the background and comoving tracks. This object remains a candidate companion and we label it as TrES-1 cc1.

The next closest object has $\Delta \Ksfilt = 6.67 \pm 0.06, \rho = 4\arcsec.940 \pm 0\arcsec.002$ and PA $=148\deg.15 \pm 0\deg.02$. \citet{Faedi2013} found a companion consistent with this detection. We include this previous measurement with our two epochs and find that the positions are consistent with both a comoving and background track. Our study shows the companion color is consistent with a late type main sequence star at the same distance as the primary star. We label this candidate companion as TreS-1 cc2.

The furthest object has $\Delta \Ksfilt = 5.7 \pm 0.1, \rho = 6\arcsec.355 \pm 0\arcsec.002$ and PA $=47\deg.31 \pm 0\deg.02$. \citet{Faedi2013} found a companion consistent with this detection. We include this previous measurement with our two epochs and find that the positions are consistent with the background track only. Therefore, we do not include this object in further analysis and we label it TrES-1 bg.

Finally, this system was also imaged by \citet{Daemgen2009}, but they did not report any companions to TrES-1. They restricted their search to companions within 2\arcsec, which would miss all three objects discussed here. In our multiplicity analysis, we count this as a candidate multi-stellar system.

\subsubsection{WASP-33}
We find a companion with $\Delta \Ksfilt = 6.11 \pm 0.02, \rho = 1\arcsec.940 \pm 0\arcsec.002$ and PA $=276\deg.25 \pm 0\deg.05$. \citet{Moya2011} finds a companion at a consistent position angle but at a separation of $1\arcsec.961 \pm 0\arcsec.003$, which is $6\sigma$ larger than our measurement. However, they report applying a rotation correction but not a NIRC2 distortion correction. Our mass and temperature estimates are also consistent with their mass (between 0.1\unit{\Msun} and 0.2\unit{\Msun}) and temperature ($3050\unit{K}\,\pm\,250\unit{K}$) estimates. They also show that the candidate companion and primary star lie on the same isochrone and argue that these objects are bound. \citet{Adams2013} also found a companion but do not report astrometric uncertainties. \citet{Woellert2015b} also report finding a companion at a position consistent with our three measurements. The separations measured over our 3 epochs are consistent with both a common proper motion track and a background track. However, the position angle measurements from this work and \citet{Moya2011} are inconsistent with a background track. With this astrometric evidence and colors and temperatures consistent with a late type main sequence star, we consider WASP-33 to be a candidate multi-stellar system.

\subsubsection{WASP-36}
We find a companion with $\Delta \Ksfilt = 2.7 \pm 0.1, \rho = 4\arcsec.869 \pm 0\arcsec.002$ and PA $=66\deg.98 \pm 0\deg.02$. This candidate companion was reported in \citet{Woellert2015b} and \citet{Evans2016}. We obtained an additional epoch in 2016. All measurements are consistent with each other and also with the background track. We expect that another epoch of Keck imaging in the next 1-2 years should allow us to determine whether or not the companion is bound.

\subsubsection{WASP-48}
We find a companion with $\Delta \Ksfilt = 7.3 \pm 0.1, \rho = 3\arcsec.571 \pm 0\arcsec.003$ and PA $=208\deg.32 \pm 0\deg.04$. Our astrometric measurements are not well-matched by models for either a bound companion or a background object. For now, we consider this a candidate multi-stellar system because the companion's color is consistent with a late type main sequence star.

This system was also imaged by \citet{Woellert2015a} but they did not report a companion. They only report detection limits out to 2\arcsec, which was at $\Delta \zpfilt =6.1$ for this target. Our temperature estimate indicates the companion is an early M dwarf, therefore, this candidate may have been below the detection limit of these previous observations.

\subsubsection{WASP-103}
We find a companion with $\Delta \Ksfilt = 1.97 \pm 0.02, \rho = 0\arcsec.239 \pm 0\arcsec.002$ and PA $=131\deg.3 \pm 0\deg.4$. The position measurements from our study and \citet{Woellert2015b} are consistent with each other, but the large uncertainty from the previous study prevents us from ruling out a background object. In addition, our companion color and effective temperature estimates are consistent with a late type main sequence star at the same distance. \citet{Evans2016} also imaged this system but reported that this companion is below their survey sensitivity.

\section{Survey Results}
\label{sec:surveyresults}
\begin{figure}
\epsscale{1.2}
\plotone{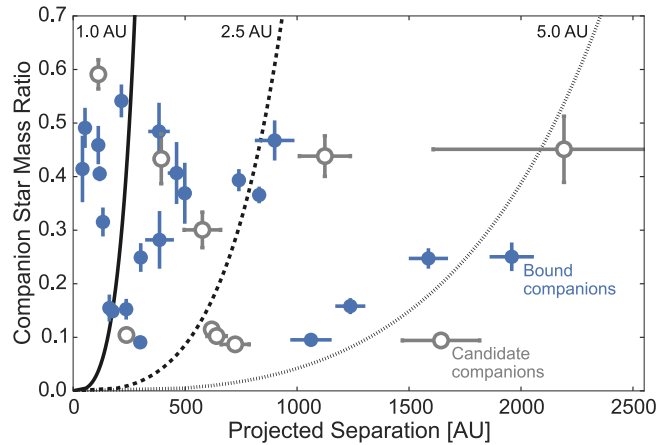}
\caption{For each companion in our survey sample, we plot the companion's mass and projected separation. Each point represents the weighted average from all observations in Table~\ref{tab:interp_sec}. The lines represent the minimum companion mass necessary to excite Kozai-Lidov oscillations at a timescale short enough to overcome general relativity pericenter precession. These representative lines assume a primary stellar mass of $1.0\unit{\Msun}$, a planetary mass of $1.0\unit{\Mjup}$, a circular planetary orbit and a stellar companion eccentricity of 0.5. The three lines (solid, dashed, and dotted) represent the difference in pericenter precession timescales for a hot Jupiter starting at $1\unit{AU}$, $2.5\unit{AU}$ and $5\unit{AU}$, respectively. Companions must be above and to the left of these lines to overcome general relativity pericenter precession timescales.
\label{fig:sep_v_mass}}
\end{figure}

We combine our new companion search sample of 27 systems (see Section~\ref{sec:obs} for a description of the sample selection) with the original sample of 50 systems surveyed in \citet{Ngo2015} in order to derive an updated estimate of the stellar multiplicity of these stars. We include all confirmed and candidate multi-stellar systems in this analysis. Although we reserve the label of confirmed companion for systems where we can demonstrate that the companion has the same proper motion as the primary, we expect that most if not all of our candidate companions are also likely to be bound. We base this argument on the fact that they have colors consistent with those of a bound companion, and also that their projected separations and contrast ratios make them unlikely to be a background object~\citep[e.g. see][]{Ngo2015,Bowler2015}.For some candidate companions, \citet{Evans2016} have suggested that a background red giant star at a moderate distance would have photometric and astrometric measurements consistent with both background and bound object tracks. Additional measurements would help to distinguish these two cases. We report a total raw stellar companion fraction of 27 out of 77 stars, or $35\,\%\,\pm\,7\,\%$. Figure~\ref{fig:sep_v_mass} shows the distribution of projected separations and mass ratios for the confirmed and candidate companions from this study and \citet{Ngo2015}.

\subsection{AO survey incompleteness correction}
\label{sec:incomplete}
We correct our raw companion fraction for survey completeness following the procedure described in~\citet{Ngo2015} for each of our 77 targets. In brief, we generate 2.5 million simulated companions over a 50x50 grid in mass and semi-major axis. Each simulated companion has an orbital eccentricity drawn from a uniform distribution~\citep{Raghavan2010} and randomized orbital elements. If the simulated companion's brightness ratio is above the $5\sigma$ contrast limit as computed in Section~\ref{sec:cc} at the projected on-sky separation, then we count it as a detection. We then calculate the average sensitivity over all grid cells where we weight each cell according to the probability that a field star would have a companion in the stated mass and semi-major axis range according to \citet{Raghavan2010}. The $i$-th target's survey sensitivity is called $S_i$ and it represents the fraction of stellar companions between $50\unit{AU}$ and $2000\unit{AU}$ (our survey phase space) that our observations could have detected. 

Next, we can use our estimate of survey completeness for each star, $S_i$, to compute the true companion fraction, $\eta$, for any arbitrary set of stars in our survey sample. We write the likelihood $L$ of observing $N_d$ detected companions out of a set of $N$ stars as:
\begin{equation}
L = \prod_{i=1}^{N_d} (S_i \eta)  \prod_{j=1}^{N-N_d}(1-S_j\eta)
\label{eqn:likelihood}
\end{equation}
where the product sum over $i$ is for the targets with a detected companion while the product sum over $j$ is for the targets without a detected companion. We define the companion fraction $\eta$ as the fraction of stars with one or more stellar companions in our survey phase space. Thus, we also make the assumption that $S_i=1$ for all systems with at least one detected companion. This is equivalent to assuming that there are no further companions within our survey phase space around targets with at least one companion already detected. This assumption is supported by our observational results and previous studies such as~\citet{Eggleton2007}.

We determine the posterior probability distribution of $\eta$ by maximizing the above likelihood via the Affine-Invariant Markov Chain Monte Carlo scheme implemented by the ``emcee'' python package~\citep{Goodman2010,ForemanMackey2013}. We use a uniform prior on $\eta$ between the possible values of $\eta=0$ and $\eta=1$. We report the $68\%$ confidence interval on $\eta$ as the uncertainties on our best estimate of $\eta$ for each of the following set of targets in our survey sample. For more details on our calculation of $S_i$, $L$, and $\eta$, see~\citet{Ngo2015}.

\subsection{Stellar companion fraction for\\ hot Jupiter hosts vs. field stars}
\begin{figure}
\epsscale{1.2}
\plotone{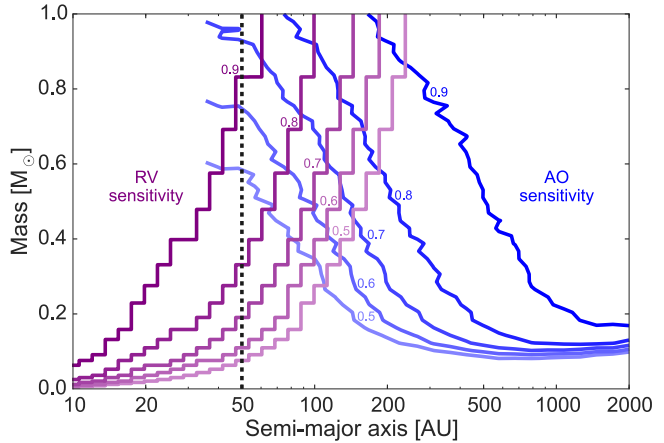}
\caption{Contours of overall sensitivity to stellar companions from long term radial velocity surveys (purple) and our AO survey (blue). These sensitivities are averaged over all targets and computed for a typical 1\unit{\Msun} target. The dashed line marks a semi-major axis of 50\unit{AU}.
\label{fig:sens}}
\end{figure}

First, we report the companion fraction of the entire survey sample to be $47\%\,\pm\,7\%$ ($47\%\,\pm\,12\%$ for the new targets presented in this work) for companions with separations between 50\unit{AU} and 2000\unit{AU}. This overall companion fraction is consistent with our previously reported companion fraction of $49\%\,\pm9\%$ in \citet{Ngo2015}. We next use the results of our long term radial velocity monitoring survey~\citep{Knutson2014,Bryan2016} to constrain the population of stellar companions within 50\unit{AU}. Following the procedure in Section 3.5 of \citet{Bryan2016}, we compute the sensitivity to stellar companions (masses greater than 0.08\unit{\Msun}) for the 50 targets in our sample with long term radial velocity data. Figure~\ref{fig:sens} shows the resulting average sensitivity contours for AO imaging and radial velocity data sets as a function of companion semi-major axis. With the exception of one target, our radial velocity monitoring rules out stellar companions within 50\unit{AU}. The only exception is the stellar companion to HAT-P-10, detected by both our radial velocity survey~\citep{Knutson2014} and our AO survey~\citep{Ngo2015} with a projected separation of 42\unit{AU}. Although the current data for this companion are also consistent with orbital semi-major axes beyond 50\unit{AU}, we count it as interior to 50\unit{AU} for the purposes of our statistical analysis. Following the same completeness-correction procedure as for our AO companion fraction, we use the RV sensitivity curves of a sample of 51 transiting hot Jupiters~\citep{Knutson2014,Bryan2016} and find that $3.9^{+4.5}_{-2.0}\%$ of these hot Jupiters have stellar companions between 1\unit{AU} and 50\unit{AU}. 

We also compare our overall companion fraction for hot Jupiter host stars with that of solar-type field stars. In \citet{Ngo2015}, we were sensitive to stellar companions with periods as short as $10^{4}\unit{days}$ for some of our nearby targets, which corresponds to separations of 10\unit{AU}. Without a constraint on potential stellar companions within 50\unit{AU} from radial velocity monitoring, we made the conservative choice to compare our AO detected companion fraction to the field star population with periods between $10^{4}\unit{days}$ and $10^{7.5}\unit{days}$ (corresponding to separations between 10\unit{AU} and 2000\unit{AU} for solar-like stars). However, surveys of star-forming regions indicate that binaries with separations less than 50\unit{AU} have significantly shorter disk lifetimes while binaries with larger separations appear to have disk lifetimes comparable to those of single stars~\citep[e.g.][]{Kraus2012}. In addition, \citet{Kraus2016arxiv} surveyed 382 {\it Kepler} planet host stars and found that there is a $4.6\sigma$ deficit in stars with binaries closer than 50\unit{AU} compared to field stars, suggesting that these close binaries negatively influence planet formation~\cite[see also][]{Wang2015b}. We therefore change our approach in this analysis to consider the multiplicity rate for companions interior and exterior to 50\unit{AU} separately.

We compute the field star companion fraction for companions with periods between $10^{5}\unit{days}$ and $10^{7.5}\unit{days}$ (corresponding to separations between 50\unit{AU} and 2000\unit{AU} for solar-like stars) to be $16\%\,\pm\,1\%$. Thus, we find that hot Jupiters have 2.9 times as many companions in this phase space as field stars, where the difference is significant at the $4.4\sigma$ level. In contrast, there is a lack of stellar companions to transiting hot Jupiter host stars with separations less than $50\unit{AU}$. On the other hand, only $3.9^{+4.5}_{-2.0}\%$ of hot Jupiters have stellar companions with separations between 1\unit{AU} and 50\unit{AU}, while $16.4\%\,\pm\,0.7\%$ of field stars have stellar companions in this range, corresponding to a $2.7\sigma$ difference. We choose a lower limit of $1\unit{AU}$ to avoid systems where the stellar companion could eject the hot Jupiter~\citep{Mardling2001,Petrovich2015b}. We note that if we relax this lower limit and considered all companions with separations less than 50\unit{AU}, we find that hot Jupiter hosts have a companion fraction of $3.9^{+4.6}_{-2.0}\%$ while field stars have a companion fraction of $22\%\,\pm\,1\%$, which is a difference of $3.8\sigma$. These values are consistent with the results of \citet{Kraus2016arxiv}.

In a recent study, \citet{Evans2016} use a sample of 101 systems observed with lucky imaging to derive a completeness-corrected estimate of $38^{+17}_{-13}\%$ for the multiplicity rate of hot Jupiter host stars. This number is in good agreement with our value, but \citet{Evans2016} differ from our study in their calculation of the equivalent field star multiplicity rate. Although their imaging survey is only sensitive to companions beyond 200 AU, they integrate over field star binaries with separations greater than 5 AU, resulting in a field star multiplicity rate of $ 35\%\,\pm\,2\% $. However, we argued above, this conflates two regions with apparently distinct companion occurrence rates. If we instead take 200\unit{AU}, or periods of $10^{5.9}\unit{days}$, as our lower limit for field star binaries and re-calculate the corresponding field star multiplicity rate we find a value of $15\%\,\pm\,1\%$, which is $1.8\,\sigma$ lower than the hot Jupiter multiplicity rate reported by \citet{Evans2016}. We therefore conclude that their results are consistent with our finding that hot Jupiters a higher multiplicity rate than field stars at wide separations. In order to facilitate comparisons between our study and those of \citet{Evans2016} and \citet{Wang2015b}, we re-calculate our hot Jupiter companion fraction for separations between 200\unit{AU}and 2000\unit{AU}.  We find a value of $32\%\,\pm\,6\%$ in this regime, in good agreement with both of these studies. This companion fraction is also $3.8\sigma$ higher than the field star companion fraction of $9.0\%\,\pm\,0.4\%$ for companions separated between $200\unit{AU}$ and $2000\unit{AU}$.

\subsection{Distribution of companion mass ratios and semi-major axes}
\label{sec:mratio}

\begin{figure}[t]
\epsscale{1.2}
\plotone{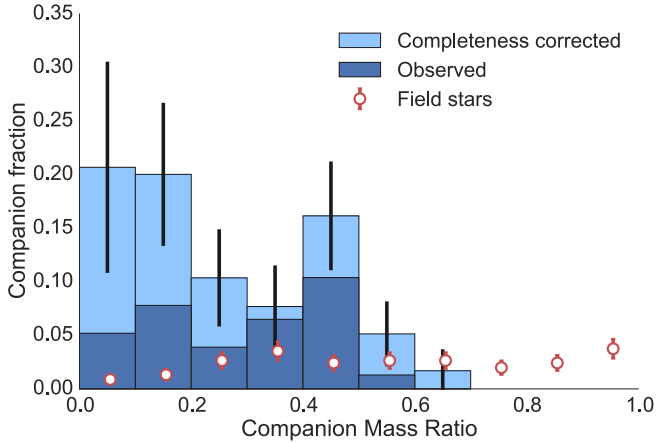}
\caption{Companion fraction as a function of companion mass ratio for targets in the completeness corrected survey sample (light blue), the uncorrected survey sample (dark blue), and the field star sample (open red symbols). The field star values (open red circles) are from \citet{Raghavan2010} and are also completeness corrected.
\label{fig:eta_mratio}}
\end{figure}

Next, we compare the observed distribution of companion mass ratios and semi-major axes with those of field stars. Figure~\ref{fig:eta_mratio} shows the survey's observed companion fraction, the survey's completeness corrected companion fraction $\eta_M$, and the completeness corrected field star companion fraction~\citep{Raghavan2010} as a function of companion star mass ratio. We find that distribution of mass ratios for the stellar companions detected in our survey is concentrated towards small values, unlike the relatively uniform distribution observed for field stars. It is possible that our distribution is shaped at least in part by observational biases in ground-based transit surveys, where binary companions with separations less than 1\arcsec\ are likely to be blended with the primary in the survey photometry, therefore diluting the observed transit depths in these systems.  Equal mass binaries with projected separations of less than 0\arcsec.5 are also challenging targets for radial velocity follow-up due to the blended nature of the stellar lines, and it is possible that these kinds of systems might receive a lower priority for follow-up as compared to apparently single stars or those with relatively faint companions. \citet{Wang2015b} found three stellar companions to {\it Kepler} short-period ($P<10\unit{days}$) giant planet hosts with $\Delta K \lesssim 0.5$, corresponding to mass ratios greater than 0.8. While this is consistent with the idea that ground-based transit surveys might be biased against detecting hot Jupiters orbiting equal mass binaries, the current transiting sample are missing this population of hot Jupiters, the current sample sizes are too small to apply a correction.

While the field star companion fraction rises up to mass ratios of 0.3, our survey companion fraction is largest for mass ratios less than 0.2. Although \citet{Raghavan2010} corrected their field star sample to account for survey incompleteness at the lowest mass ratios, it is possible that their correction underestimated the true incompleteness at small mass ratios. Because this trend is seen in the completeness corrected companion fraction but not the observed companion fraction, we considered whether it could be an artifact introduced by our completeness correction calculation. We generate simulated companions down to a mass of 0.08\unit{\Msun}, which is a mass ratio of 0.05 for our most massive survey target and less than 0.1 for all but one of our survey targets (for WASP-43, this limit corresponds to a mass ratio of 0.13). Therefore, while the smallest mass ratio bin may have unequal sizes for each target, the second smallest bin is the same for all targets and also shows an enhanced companion fraction relative to that of field stars. Although our correction is more uncertain at lower masses, the difference between our completeness corrected companion fraction and the field star distribution in the 0.1-0.2 mass ratio bin is greater than the uncertainty by $2.8\sigma$.

\begin{figure}[t]
\epsscale{1.2}
\plotone{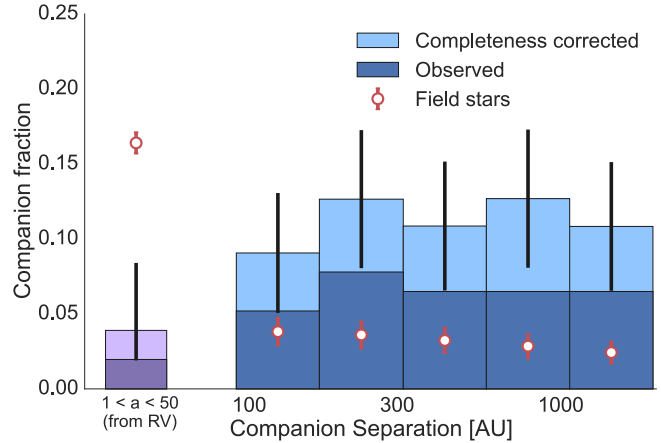}
\caption{Companion fraction as a function of companion separation, in logarithmic bins, for targets in the completeness corrected survey sample (light blue), the uncorrected survey sample (dark blue), and the field star sample (open red symbols). The leftmost bin represents the corrected (light purple) and uncorrected (dark purple) companion fraction from 1\unit{AU} to 50\unit{AU}, computed from long term RV sensitivity surveys. The field star values (open red circles) are from \citet{Raghavan2010} and are also completeness corrected.
\label{fig:eta_logsepAU}}
\end{figure}

Figure~\ref{fig:eta_logsepAU} shows the survey's observed companion fraction, the survey's completeness corrected companion fraction $\eta_S$, and the completeness corrected field star companion fraction~\citep{Raghavan2010} as a function of companion star projected separation. The comparison is made in logarithmic space for projected separation as~\citet{Raghavan2010} found that the periods of companion stars follow a log-normal distribution. This plot shows a higher companion fraction in our survey than in the field star sample. However, we find that the relative distribution of companion separations in our sample is in good agreement with those of the field star sample. Although our distribution appears to be effectively uniform, this is consistent with the log-normal distribution reported in~\citet{Raghavan2010} since our survey space spans a relatively small fraction of the separations considered in \citet{Raghavan2010}. 

\subsection{Multiplicity and host star metallicity}
We next investigate whether our measured companion fraction could be affected by differences in the metallicities of the stars in our sample as compared to the field star sample. \citet{Raghavan2010} found tentative evidence for a rise in the multiplicity rate for metal-poor ([Fe/H] $< -0.3$) stars and a uniform multiplicity rate for metallicities between -0.3 and +0.4. Our targets have metallicities ranging from $-0.29$ to $+0.45$. We therefore conclude that the increased companion fraction for our sample of hot Jupiter hosts is unlikely to be due to the higher metallicities of our stars as compared to the field star sample.

We also considered whether the presence of companions in our sample is correlated with the metallicities of the host stars, although we would not expect such a correlation based on the results from the field star sample. If we simply compare the host star metallicity distribution of single and multi-stellar systems, we find that they are consistent with each other. This is not surprising, as the typical metallicity uncertainties are between 0.1 dex and 0.2 dex, which is a significant fraction of the total metallicity range spanned by our sample.

\section{Discussion}
\label{sec:discuss}
Our survey results show that stellar companions are found in hot Jupiter systems at a rate which is higher than the rate for field stars, that these companions tend to have low mass ratios, and that their distribution of projected separations is similar to that of field stars over the range of separations considered here (50\unit{AU} to 2000\unit{AU}). Here, we discuss two potential ways in which companion stars might influence hot Jupiter formation. We first consider whether these wide stellar companions could enhance the global gas giant planet formation rate, and then consider whether or not they might preferentially enable the inward migration of gas giant planets formed at larger separations.

\subsection{Are multi-stellar systems more favorable for gas giant planet formation?}
\label{sec:moreGiants}
One possible explanation for the higher multiplicity rate of hot Jupiter host stars is that these systems are more favorable sites for gas giant planet formation than single stars. For example, a stellar companion could raise spiral arms in a protoplanetary disk. These spiral arms are regions of high particle and gas density, which may be conducive to giant planet formation~\citep[e.g. dust traps as in][]{vanderMarel2013}. Indeed, planetesimal formation through the streaming instability~\citep{Youdin2005,Johansen2007b} as well as subsequent core growth through pebble accretion~\citep{Lambrechts2014a,Lambrechts2014b} exhibit a strong dependence on the local density of solids~\citep{Carrera2015}. Recent high contrast VLT/SPHERE imaging of the protoplanetary disk around HD 100453, which has an M dwarf companion located at a distance of 120\unit{AU}, revealed the presence of spiral structures~\citep{Wagner2015}. \citet{Dong2015} showed that these structures are best explained as perturbations from this companion rather than processes intrinsic to the disk. HD 141569 is part of a triple system and also hosts an asymmetric disk~\citep[for a summary of these features see][and references therein]{Biller2015} with a structure that can plausibly be attributed to perturbations from these stellar companions~\citep{Augereau2004,Quillen2005}. The mass ratios and separations of these two systems are similar to those of the binaries in our study, suggesting that the presence of a stellar companion can facilitate planet formation in these systems.

Alternatively, protoplanetary disks around wide binaries might be more massive than those around single stars, and therefore would have more material available for giant planet formation.  Although current observations suggest that close ($<50\unit{AU}$ separation) binaries have shorter disk lifetimes, disks in wide binaries appear to have lifetimes comparable to those of isolated stars~\citep{Kraus2012}. Planet formation simulations predict that higher-mass disks will form higher-mass planets~\citep[e.g.][]{Thommes2008,Mordasini2012}. In addition, \citet{Duchene2013} suggest that the timing of fragmentation in protostellar disks could create an asymmetric mass distribution resulting in a low mass ratio companion with a relatively small disk as compared to that of the primary star. These scenarios assume that the companion stars formed at the same time as the primary star, rather than being captured. 

High contrast imaging and radial velocity surveys of planet-hosting stars in the {\it Kepler} sample suggest that binary star systems are less likely to host small, close-in planets than their single counterparts~\citep{Wang2014}.  Although this might be interpreted as an argument against the massive disk scenario, it might conversely be argued that rocky cores embedded in a more massive disk are more likely to reach runaway accretion and turn into gas giants~\citep[e.g.][]{Ikoma2000,Lee2015}. These gas giant planets could then become hot Jupiters via Type II disk migration~\citep[e.g.][]{Lin1996} or via interactions with the stellar companion as described below. Additionally, cores that reside in close proximity to the host star may also undergo runaway accretion, leading to in-situ formation of hot Jupiters~\citep{Bodenheimer2000,Batygin2015arxiv,Boley2016}.

\subsection{Are binary systems causing planets to migrate inwards via Kozai-Lidov oscillations?}
\label{sec:maxKL}
We next consider a scenario in which gas giants form at the same rate around both single and binary stars, but the presence of a stellar companion causes these planets to migrate inward from their formation locations via three-body interactions such as Kozai-Lidov oscillations~\citep[e.g.][]{Fabrycky2007,Naoz2012}. We compute representative minimum mass ratios as a function of companion separation required for the stellar companion to excite Kozai-Lidov oscillations on a $1\unit{\Mjup}$ mass planet. In single planet systems, this constraint is set by the planet pericenter precession timescale caused by general relativity. We therefore calculate the companion mass and separation such that the Kozai-Lidov oscillation timescale is equal to the pericenter precession timescale, following Equations 1 and 23 in \citet{Fabrycky2007}. For these representative limits, we assume a primary star mass of $1\unit{\Msun}$, a companion star orbital eccentricity of 0.5 and a circular orbit for the planet. These expressions scale with the companion star's orbital eccentricity as $(1-e^2)^{1/2}$ and with the planet's orbital eccentricity as $(1-e^2)^{-1/3}$, so the effect of a non-zero planetary eccentricity is mild. We choose 0.5 as the representative stellar eccentricity as previous studies of stellar companions around FGK stars in our solar neighborhood show that stellar companions with periods longer than 12 days have eccentricities uniformly distributed between 0 and 1~\citep{Raghavan2010}. We compute three different representative limits for planets with starting semimajor axis distances of $1\unit{AU}$, $2.5\unit{AU}$ and $5\unit{AU}$, and compare these limits to the masses and projected separations of our observed population of stellar companions in Figure~\ref{fig:sep_v_mass}.

We next compute the completeness corrected fraction of hot Jupiter systems with stellar companions that are capable of inducing Kozai-Lidov oscillations. Unlike the calculation of representative cases above, we now use actual system parameters for each target, including the primary star mass and planet mass. Unfortunately the orbital parameters of the companion star, such as eccentricity and inclination, are unknown because our baselines are not currently long enough to detect orbital motion in these systems. Because the Kozai-Lidov timescale depends only weakly on the eccentricity of the companion for values less than 0.9, we obtain equivalent results if we set the eccentricities of the companions to 0.5 in our distributions as compared to sampling from a uniform distribution. We assume that if a companion is present, its mutual inclination will be greater than the critical angle required to induce Kozai-Lidov oscillations. We do not account for suppression of the stellar Kozai-Lidov due to interactions with other planetary or brown dwarf companions, which are known to exist in a subset of these systems~\citep{Wu2003,Batygin2011,Knutson2014}. 

The resulting numbers therefore represent an upper limit on the fraction of hot Jupiter systems that have experienced Kozai-Lidov in the most optimistic case. We compute these fractions for three different initial planetary semimajor axes, at 1\unit{AU}, 2.5\unit{AU}, and 5.0\unit{AU}, and we find that the upper limits to be $16\%\,\pm\,6\%$, $34\%\,\pm\,7\%$, and $47\%\,\pm\,7\%$, respectively. We also average over all potential initial planetary semimajor axes between 1\unit{AU} to 5\unit{AU} by sampling from the \citet{Cumming2008} power law distribution fit to the population of known RV-detected gas giant planets. We find that the upper limit on the fraction of hot Jupiter systems that formed via Kozai-Lidov migration in this case is $32\%\,\pm7\%$. 

We also consider a more realistic case in which we account for the fact that the presence of additional gas giant planetary companions would act to disrupt Kozai-Lidov oscillations~\citep{Wu2003,Batygin2011}. \citet{Knutson2014} found that $51\% \pm 10\%$ of hot Jupiter systems have long period RV-detected companions so we multiply our optimistic Kozai-Lidov upper limits by a factor of 0.49 and find that our realistic upper limit on the fraction of hot Jupiter systems that formed via Kozai-Lidov migration is $16\%\,\pm5\%$. Although a critical mutual inclination, which depends on the planet's initial eccentricity, is required for this mechanism, we do not know the stellar companion inclination distribution for hot Jupiter systems or the eccentricity distribution of proto-hot Jupiters. If we assume an isotropic distribution of stellar companions, then our corresponding upper limit on the fraction of Kozai-Lidov systems will decrease by a factor of 0.37, to $6\%\,\pm\,2\%$. However, if Kozai-Lidov migration is a strong contributor to hot Jupiter migration, then it is possible that the inclination distribution for hot Jupiter companions are not isotropic. In addition, \citet{Martin2016} show that planet-disc interactions in binary star systems can act to tilt the planet's orbit so that the angle between the planet and companion is greater than the critical angle. We therefore conclude that the inclusion of a geometric correction for companion inclination is not currently justified, leaving us with an estimate of $16\%\,\pm\,5\%$ for the fraction of hot Jupiters that might have migrated via Kozai-Lidov oscillations induced by a stellar companion.

These upper limits are consistent with a range of recent theoretical work constraining the frequency of Kozai-Lidov oscillations in hot Jupiter systems. Simulations of binary star planet hosting systems considering the eccentric Kozai-Lidov mechanism to octopole order find that the eccentric Kozai-Lidov mechanism can account for the formation of up to 30\% of hot Jupiter systems~\citep{Naoz2012}. \citet{Dawson2015} estimate the $2\sigma$ upper limit on the fraction of hot Jupiters with periods greater than 3 days that could have migrated in via Kozai-Lidov interactions with a stellar companion to be 44\%, based on the relatively long circularization timescales in these systems and the corresponding absence of a large population of high eccentricity gas giants at intermediate separations in the Kepler sample. This calculation implicitly assumes that all systems have an outer planetary or stellar companion capable of inducing a high eccentricity in the proto-hot Jupiter, but does not specifically require that this occur via Kozai-Lidov oscillations. \citet{Petrovich2015a} performed simulations similar to \citet{Naoz2012} with a more restrictive value for the tidal disruption distance, that is, the pericenter distance where an inwardly migrating planet would be tidally disrupted instead of forming a hot Jupiter. When he considers the currently observed hot Jupiter occurrence rate and semi-major axis distribution, he finds that at most 23\% of observed hot Jupiters could have been formed via Kozai-Lidov migration. We note that both \citet{Naoz2012} and \citet{Petrovich2015a} assumed that hot Jupiter host stars have companions at the same rate as field stars, which means their limits are underestimated by a factor of two. However, they also assume that the proto-hot Jupiter is the only planet in the system, resulting in a factor of two over-estimate which effectively cancels the under-estimate due to the enhanced binary rate in these systems. \cite{Anderson2016} and \citet{Munoz2016} performed an analytical calculation of the fraction of hot Jupiters created via a Kozai-Lidov migration scenario and found values ranging from 12\% to 15\%, depending on initial planet masses from 0.3 to 3 Jupiter masses and varying tidal dissipation strength.

As demonstrated in Figure~\ref{fig:sep_v_mass}, the upper limit we derive here is primarily sensitive to our assumptions about the starting semi-major axes of the proto-hot Jupiters. Because the stellar companions detected in our survey typically have low masses and large projected separations, many of them require large initial semi-major axes for the planet in order to achieve the required inward migration. In addition, if hot Jupiter survey selection biases exclude hot Jupiters in equal mass binaries (see Section~\ref{sec:mratio}), then our sample may not be representative of the entire population of hot Jupiters. Nevertheless, Kozai-Lidov oscillations cannot be the dominant migration mechanism for transiting hot Jupiter systems from ground-based surveys.

It is worth noting that there are other ways in which a stellar companion might affect the dynamical evolution of planetary systems beyond the Kozai-Lidov migration scenario considered here. For example, \citet{Batygin2012} and \citet{Spalding2014,Spalding2015} have proposed that the presence of a companion could change the orientation of the protoplanetary disk relative to the star's spin axis. It is our hope that the observations described here will serve to motivate new studies of the effects of the observed population of stellar companions on the dynamical evolution of these systems.  We expect that future observations, e.g. by {\it Gaia}, may also provide additional constraints on the orbital properties of these stellar companions, at least in the subset of systems for which it is possible to detect astrometric motion of the secondary on several year timescales.

\section{Summary}
\label{sec:summary}
We conducted a direct imaging search for stellar companions around 77 transiting gas giant planet hosts and combine our results with a radial velocity stellar companion surveys to determine the occurrence of stellar companions around hot Jupiter hosts. We detected a total of 27 candidate stellar companions, including three companions reported for the first time in this study. We also followed up on five systems with known candidate companions identified in published surveys. For all detected companions, we measure their flux ratios and positions to characterize the companion properties and evaluate the likelihood that they are physically bound to their host stars. We also provide updated photometric and astrometric measurements for all systems, including previously published candidate companions. We confirm common proper motion for six new multi-stellar systems while the other nine examined in this study remain candidate multi-stellar systems.

Overall, we find that hot Jupiters have a stellar companion rate of $47\%\,\pm\,7\%$ for companions between 50\unit{AU} and 2000\unit{AU}. This is $4.4\sigma$ larger than the equivalent companion rate for field stars, which is $16\%\,\pm\,1\%$. For companions between 1\unit{AU} and 50\unit{AU} we find that only $3.9^{+4.5}_{-2.0}\%$ of hot Jupiter systems host stellar companions while field stars have a companion rate of $16.4\%\,\pm\,0.7\%$, corresponding to a difference of $2.7\sigma$. We suggest that there may be a connection between the presence of a companion star beyond 50\unit{AU} and processes that either favor giant planet formation or facilitate the inward migration of planets in these systems. 

We examine the companion fraction as a function of companion mass and companion separation and compare these distributions to those of field star binaries. We find that the mass ratio distribution for binaries hosting hot Jupiters peaks at small mass ratios, unlike the relatively uniform distribution of mass ratios observed for field star binaries. Although this may in part reflect a bias against equal mass binaries in photometric transit surveys, it is also plausible that higher mass companions might actively suppress planet formation by disrupting the protoplanetary disk.  As discussed in Section~\ref{sec:moreGiants}, the more subtle effects of a low-mass companion on the disk structure could instead aid in planet formation by creating regions of locally enhanced density. Alternatively, protoplanetary disk masses in binary star systems may be higher than those of their isolated counterparts, resulting in globally enhanced disk densities. We also find that the companion fraction does not depend strongly on companion separation for semi-major axes greater than 50\unit{AU}.

We additionally use our sample of resolved stellar binaries to calculate an upper limit on the fraction of hot Jupiter systems that might have migrated inward via Kozai-Lidov oscillations. We evaluate this number as a function of the planet's initial semi-major axis and find that the upper limits are $16\%\,\pm\,6\%$, $34\%\,\pm\,7\%$, and $47\%\,\pm\,7\%$, for initial semi-major axes of 1\unit{AU}, 2.5\unit{AU}, and 5\unit{AU}, respectively. When averaged over 1-5\unit{AU} using the best-fit power law distribution for RV-detected planets and accounting for the presence of radial velocity companions in a subset of the systems observed, this upper limit is $16\%\,\pm\,5\%$. These observational constraints are in good agreement with published theoretical models and simulations of hot Jupiter formation via the Kozai-Lidov mechanism, which also suggest that Kozai-Lidov driven migration can only account for a small fraction of the known hot Jupiter systems.



\acknowledgments

This work was supported by NASA grant NNX14AD24G. HN is grateful for funding support from the Natural Sciences and Engineering Research Council of Canada and the NASA Earth and Space Science Fellowship Program grant NNX15AR12H. 

This work was based on observations at the W. M. Keck Observatory granted by the California Institute of Technology. We thank the observers who contributed to the measurements reported here and acknowledge the efforts of the Keck Observatory staff. We extend special thanks to those of Hawaiian ancestry on whose sacred mountain of Mauna Kea we are privileged to be guests.

{\it Facility:} \facility{Keck:II (NIRC2)}


\bibliographystyle{apj}

\clearpage
\LongTables
\begin{deluxetable*}{lclccccc}
\tablecolumns{8}
\tabletypesize{\scriptsize}
\tablecaption{Summary of NIRC2 AO Observations  \label{tab:obs}}
\tablewidth{0pt}
\tablehead{
\colhead{Target} & \colhead{$N_{cc}$} & \colhead{UT Obs. Date} & \colhead{Filter} & \colhead{Array} & \colhead{$T_{\mathrm{int}}$} & \colhead{$N_{\mathrm{fit}}$} & \colhead{$N_{\mathrm{stack}}$}
}
\startdata
\sidehead{\bf Survey targets}
HAT-P-1\tablenotemark{a} & 0 & 2013 Oct 17 & \Kpfilt & 1024 & 9.0 & \nodata & 12 \\
HAT-P-3 & 0 & 2013 May 31 & \Ksfilt & 1024 & 9.0 & \nodata & 12 \\
HAT-P-5 & 1 & 2013 Jul 04 & \Ksfilt & 1024 & 10.0 & 4 & 12 \\
 &  & 2015 Jun 24 & \Jfilt & 1024 & 12.5 & 12 & \nodata \\
 &  & 2015 Jun 24 & \Ksfilt & 1024 & 12.5 & 12 & 12 \\
HAT-P-9 & 0 & 2013 Mar 02 & \Ksfilt & 1024 & 10.0 & \nodata & 12 \\
HAT-P-19 & 0 & 2013 Aug 19 & \Ksfilt & 1024 & 12.5 & \nodata & 12 \\
HAT-P-21 & 0 & 2013 Mar 02 & \Ksfilt & 1024 & 10.0 & \nodata & 12 \\
HAT-P-23 & 0 & 2013 Jun 22 & \Ksfilt & 1024 & 25.0 & \nodata & 12 \\
HAT-P-25 & 0 & 2014 Nov 10 & \Ksfilt & 1024 & 12.0 & \nodata & 12 \\
HAT-P-27 & 1 & 2014 Jul 12 & \Jfilt & 1024 & 15.0 & 12 & \nodata \\
 &  & 2014 Jul 12 & \Hfilt & 1024 & 15.0 & 12 & \nodata \\
 &  & 2014 Jul 12 & \Ksfilt & 1024 & 15.0 & 12 & 12 \\
 &  & 2015 Jan 09 & \Jfilt & 1024 & 12.5 & 12 & \nodata \\
 &  & 2015 Jan 09 & \Hfilt & 1024 & 12.5 & 12 & \nodata \\
 &  & 2015 Jan 09 & \Ksfilt & 1024 & 12.5 & 12 & 12 \\
 &  & 2015 Jun 24 & \Ksfilt & 1024 & 12.5 & 12 & 12 \\
HAT-P-28 & 1 & 2015 Jul 07 & \Jfilt & 1024 & 15.0 & 12 & \nodata \\
 &  & 2015 Jul 07 & \Ksfilt & 1024 & 15.0 & 12 & 12 \\
 &  & 2015 Jul 10 & \Jcfilt & 1024 & 25.0 & 6 & \nodata \\
 &  & 2015 Jul 10 & \BrGfilt & 1024 & 22.0 & 6 & 6 \\
HAT-P-29\tablenotemark{b} & 1 & 2012 Feb 02 & \Jfilt & 1024 & 10.0 & 9 & \nodata \\
 &  & 2012 Feb 02 & \Kpfilt & 1024 & 15.0 & 9 & 9 \\
 &  & 2015 Jul 05 & \Ksfilt & 1024 & 30.0 & 6 & 6 \\
 &  & 2015 Jul 10 & \BrGfilt & 1024 & 10.0 & 4 & 6 \\
HAT-P-35 & 1 & 2013 Mar 02 & \Jfilt & 1024 & 10.0 & 9 & \nodata \\
 &  & 2013 Mar 02 & \Ksfilt & 1024 & 10.0 & 12 & 12 \\
 &  & 2014 Nov 10 & \Jfilt & 1024 & 12.0 & 12 & \nodata \\
 &  & 2014 Nov 10 & \Hfilt & 1024 & 12.5 & 12 & \nodata \\
 &  & 2014 Nov 10 & \Ksfilt & 1024 & 12.0 & 12 & 12 \\
HAT-P-36 & 0 & 2013 Mar 02 & \Ksfilt & 1024 & 10.0 & \nodata & 12 \\
HAT-P-37 & 0 & 2015 Jun 24 & \Ksfilt & 1024 & 12.0 & \nodata & 12 \\
HAT-P-38 & 0 & 2015 Jul 07 & \Ksfilt & 1024 & 15.0 & \nodata & 7 \\
HAT-P-39 & 1 & 2013 Mar 02 & \Jfilt & 1024 & 10.0 & 12 & \nodata \\
 &  & 2013 Mar 02 & \Ksfilt & 1024 & 10.0 & 12 & 12 \\
 &  & 2014 Nov 07 & \Jfilt & 1024 & 10.0 & 11 & \nodata \\
 &  & 2014 Nov 07 & \Hfilt & 1024 & 10.0 & 12 & \nodata \\
 &  & 2014 Nov 07 & \Ksfilt & 1024 & 10.0 & 12 & 12 \\
HAT-P-40 & 0 & 2014 Oct 03 & \Ksfilt & 1024 & 15.0 & \nodata & 12 \\
HAT-P-41 & 1 & 2014 Oct 03 & \Ksfilt & 1024 & 12.5 & 6 & 6 \\
 &  & 2015 Jun 24 & \Jfilt & 1024 & 12.5 & 12 & \nodata \\
 &  & 2015 Jun 24 & \Ksfilt & 1024 & 12.5 & 12 & 12 \\
HAT-P-42 & 0 & 2015 Jan 10 & \Ksfilt & 1024 & 15.0 & \nodata & 12 \\
HAT-P-43 & 0 & 2014 Nov 10 & \Ksfilt & 1024 & 12.0 & \nodata & 12 \\
TrES-1 & 2 & 2013 Jul 04 & \Ksfilt & 1024 & 9.0 & 4 & 12 \\
 &  & 2015 Jun 24 & \Jfilt & 1024 & 12.5 & 12 & \nodata \\
 &  & 2015 Jun 24 & \Ksfilt & 1024 & 12.5 & 12 & 12 \\
WASP-5 & 0 & 2013 Oct 17 & \Kpfilt & 1024 & 10.0 & \nodata & 12 \\
WASP-13 & 0 & 2015 Jan 10 & \Ksfilt & 1024 & 13.6 & \nodata & 15 \\
WASP-33 & 1 & 2013 Aug 19 & \Jfilt & 256 & 9.0 & 6 & \nodata \\
 &  & 2013 Aug 19 & \Hfilt & 256 & 9.0 & 6 & \nodata \\
 &  & 2013 Aug 19 & \Ksfilt & 512 & 10.6 & 12 & 12 \\
 &  & 2014 Dec 07 & \Ksfilt & 512 & 15.0 & 12 & 12 \\
 &  & 2015 Dec 26 & \Ksfilt & 512 & 15.9 & 12 & 12 \\
WASP-39 & 0 & 2013 Jul 04 & \Ksfilt & 1024 & 10.0 & \nodata & 12 \\
WASP-43 & 0 & 2013 Mar 02 & \Ksfilt & 1024 & 10.0 & \nodata & 12 \\
WASP-48 & 1 & 2013 Aug 19 & \Ksfilt & 1024 & 12.5 & 8 & 12 \\
 &  & 2015 Jun 24 & \Jfilt & 1024 & 12.5 & 12 & \nodata \\
 &  & 2015 Jun 24 & \Ksfilt & 1024 & 12.5 & 12 & 12 \\
XO-1 & 0 & 2015 Jun 24 & \Ksfilt & 1024 & 12.5 & \nodata & 12 \\

\sidehead{\bf Non-survey targets}
HAT-P-54 & 1 & 2016 Jan 25 & \Ksfilt & 1024 & 15.0 & 12 & 12 \\
WASP-36 & 1 & 2016 Jan 25 & \Ksfilt & 1024 & 15.0 & 9 & 9 \\
WASP-58 & 1 & 2015 Jul 10 & \Jcfilt & 1024 & 18.0 & 6 & \nodata \\
 &  & 2015 Jul 10 & \BrGfilt & 1024 & 12.0 & 6 & 6 \\
WASP-76 & 1 & 2015 Jul 10 & \BrGfilt & 1024 & 1.5 & 3 & 3 \\
 &  & 2015 Jul 10 & \Jcfilt & 1024 & 1.1 & 3 & \nodata \\
WASP-103 & 1 & 2016 Jan 25 & \Jfilt & 1024 & 15.0 & 12 & \nodata \\
 &  & 2016 Jan 25 & \Ksfilt & 1024 & 15.0 & 12 & 12 \\
 &  & 2016 Jan 25 & \Hfilt & 1024 & 15.0 & 12 & \nodata
\enddata
\tablecomments{Column $N_{cc}$ is the number of candidate companions detected. Column ``Array'' is the horizontal size, in pixel, of the NIRC2 array readout region and corresponds to subarray sizes of 1024x1024 (the full NIRC2 array), 512x512, or 256x264. Column $T_{\mathrm{int}}$ is the total integration time, in seconds, of a single frame. Column $N_{\mathrm{fit}}$ is the number of frames used in our photometric and/or astrometric analysis, and is only given when companions are present. Column $N_{\mathrm{stack}}$ is the number of frames combined to make the contrast curve measurements. We only compute contrast curves in the \Kpfilt, \Ksfilt, \Kcfilt, \BrGfilt\ bandpasses so this column is not applicable for other bandpasses. In some cases, $N_{\mathrm{fit}}$ and $N_{\mathrm{stack}}$ are not equal because the companion may not be present in all frames due to the dither pattern and/or observing conditions.}
\tablenotetext{a}{HAT-P-1 has a known stellar companion~\citep{Liu2014b} with a similar mass but at a separation of 11\arcsec.3, it is outside of our survey's field of view.}
\tablenotetext{b}{We originally reported no companions around HAT-P-29 in \citet{Ngo2015}. However, \citet{Woellert2015b} reports a faint companion that we had missed earlier. We recovered this companion in our old images and also followed up with more observations in July 2015.}
\end{deluxetable*}

\clearpage
\begin{deluxetable*}{lcccccccc}
\tablecolumns{9}
\tabletypesize{\scriptsize}
\tablewidth{0pt}
\tablecaption{Target stellar parameters\label{tab:stellar_params}}
\tablehead{
& & \colhead{$T_{\mathrm{eff}}$}  & \colhead{$M$} & \colhead{$\log g$} & \colhead{$D$}& \multicolumn{3}{c}{References for...} \vspace{-0.1cm} \\
\colhead{Target} & \colhead{$N_{cc}$} & \vspace{-0.2cm}& & & & & & \\
 & & \colhead{(K)} & \colhead{(\Msun)} & \colhead{(cgs)} & \colhead{(pc)} & \colhead{$T$} & \colhead{$M,\log g$} & \colhead{$D$}
}
\startdata
\sidehead{\bf Survey targets}
HAT-P-1   & 0 & $5980 \pm 49$   & $1.151 \pm 0.052$ & $4.359 \pm 0.014$ & $155 \pm 15$ & 1 & 1 & 2\\
HAT-P-3   & 0 & $5185 \pm 80$   & $0.917 \pm 0.030$ & $4.594 \pm 0.041$ & $166.4 \pm 14.4$ & 3 & 3 & 4\\
HAT-P-5   & 1 & $5960 \pm 100$ & $1.163 \pm 0.069$ & $4.39 \pm 0.04$\tablenotemark{a}  & $340 \pm 30$ & 2 & 5 & 6\\
HAT-P-9   & 0 & $6350 \pm 150$ & $1.28 \pm 0.10$     & $4.293 \pm 0.046$ & $480 \pm 60$ & 7 & 7 & 8\\
HAT-P-19 & 0 & $4990 \pm 130$ & $0.842 \pm 0.042$ & $4.54 \pm 0.05$     & $215 \pm 15$ & 9 & 9 & 9\\
HAT-P-21 & 0 & $5588 \pm 80$   & $0.947 \pm 0.042$ & $4.33 \pm 0.06$     & $254 \pm 19$ & 10 & 10 & 10\\
HAT-P-23 & 0 & $5885 \pm 72$   & $1.104 \pm 0.047$ & $4.407 \pm 0.018$ & $355.0 \pm 40.8$ & 11 & 11 & 4\\
HAT-P-25 & 0 & $5500 \pm 80$   & $1.010 \pm 0.032$ & $4.48 \pm 0.04$     & $297^{+17}_{-13}$ & 12 & 12 & 12\\
HAT-P-27 & 1 & $5300 \pm 90$   & $0.945 \pm 0.035$ & $4.51 \pm 0.04$     & $204 \pm 14$ & 13 & 13 & 13\\
HAT-P-28 & 1 & $5680 \pm 90$   & $1.025 \pm 0.047$ & $4.36 \pm 0.06$     & $395^{+34}_{-26}$ & 14 & 14 & 14\\
HAT-P-29\tablenotemark{b} & 1 & $6086 \pm 69$ & $1.207 \pm 0.046$ & $4.34 \pm 0.06$ & $322^{+35}_{-21}$ & 15 & 15 & 14\\
HAT-P-35 & 1 & $6178 \pm 45$   & $1.16 \pm 0.08$     & $4.40 \pm 0.09$     & $535 \pm 32$ & 16 & 16 & 17 \\
HAT-P-36 & 0 & $5620 \pm 40$   & $1.030 \pm 0.042$ & $4.416 \pm 0.011$  & $317 \pm 17$ & 18 & 18 & 17\\
HAT-P-37 & 0 & $5500 \pm 100$ & $0.929 \pm 0.043$ & $4.52 \pm 0.04$\tablenotemark{a}  & $411 \pm 26$ & 17 & 17 & 17\\
HAT-P-38 & 0 & $5330 \pm 100$ & $0.886 \pm 0.044$ & $4.45^{+0.06}_{-0.07}$\tablenotemark{a} & $249^{+26}_{-19}$ & 19 & 19 & 19\\
HAT-P-39 & 1 & $6430 \pm 100$ & $1.404 \pm 0.051$ & $4.16 \pm 0.03$\tablenotemark{a}   & $641^{+115}_{-66}$ & 20 & 20 & 20\\
HAT-P-40 & 0 & $6080 \pm 100$ & $1.512 \pm 0.109$ & $3.93 \pm 0.01$\tablenotemark{a}   & $548 \pm 36$ & 20 & 20 & 20\\
HAT-P-41 & 1 & $6479 \pm 51$   & $1.28 \pm 0.09$     & $4.39 \pm 0.22$      & $311^{+36}_{-27}$ & 21 & 21 & 20\\
HAT-P-42 & 0 & $5743 \pm 50$   & $1.178 \pm 0.068$ & $4.14 \pm 0.07$      & $414 \pm 51$ & 22 & 22 & 22\\
HAT-P-43 & 0 & $5645 \pm 74$   & $1.048 \pm 0.042$ & $4.37 \pm 0.02$      & $566^{+67}_{-37}$ & 22 & 22 & 22\\
TrES-1     & 2 & $5226 \pm 38$   & $0.85 \pm 0.07$     & $4.40 \pm 0.10$      & $129.7 \pm 8.7$ & 16 & 16 & 4\\
WASP-5   & 0 & $5785 \pm 83$   & $1.00 \pm 0.08$     & $4.54 \pm 0.14$      & $318.6 \pm 19.9$ & 16 & 16 & 4\\
WASP-13 & 0 & $6025 \pm 21$   & $1.20 \pm 0.08$     & $4.19 \pm 0.03$      & $155 \pm 18$ & 16 & 16 & 23\\
WASP-33 & 1 & $7430 \pm 100$ & $1.495 \pm 0.031$ & $4.3 \pm 0.2$          & $123.1 \pm 7.2$ & 24 & 24 & 4\\
WASP-39 & 0 & $5400 \pm 150$ & $0.93 \pm 0.034$   & $4.50 \pm 0.01$\tablenotemark{a}    & $230 \pm 80$ & 25 & 25 & 25\\
WASP-43 & 0 & $4400 \pm 200$ & $0.58 \pm 0.05$     & $4.64 \pm 0.02$\tablenotemark{a}    & $106.1 \pm 7.2$ & 26 & 26 & 4\\
WASP-48 & 1 & $6000 \pm 150$ & $1.062 \pm 0.075$ & $4.101 \pm 0.023$   & $466.0 \pm 49.0$ & 11 & 11 & 4\\
XO-1        & 0 & $5754 \pm 42$   & $0.93 \pm 0.07$     & $4.61 \pm 0.05$       & $177.9 \pm 10.7$ & 16 & 16 & 4\\
\sidehead{\bf Non-survey targets}
HAT-P-54 & 1 & $4390 \pm 50$   & $0.645 \pm 0.020$ & $4.667 \pm 0.012$   & $135.8 \pm 3.5$ & 27 & 27 & 27 \\
WASP-36 & 1 & $5928 \pm 59$   & $1.00 \pm 0.07$     & $4.51 \pm 0.09$       & $450 \pm 120$ & 16 & 16 & 28\\
WASP-58 & 1 & $5800 \pm 150$ & $0.94 \pm 0.1$       & $4.27 \pm 0.09$       & $300 \pm 50$ & 29 & 29 & 29\\
WASP-76 & 1 & $6250 \pm 100$ & $1.46 \pm 0.07$     & $4.128 \pm 0.015$   & $120 \pm 20$ & 30 & 30 & 30\\
WASP-103 & 1 & $6110 \pm 160$ & $1.220^{+0.039}_{-0.036}$ & $4.22^{+0.12}_{-0.05}$ & $470 \pm 35$ & 31 & 31 & 31
\enddata
\tablecomments{$N_{cc}$ is the number of candidate companions detected.}
\tablerefs{(1) \citet{Nikolov2014}; (2) \citet{Torres2008}; (3) \citet{Chan2011}; (4) \citet{Triaud2014}; (5) \citet{Southworth2012b}; (6) \citet{Bakos2007c}; (7) \citet{Southworth2012c}; (8) \citet{Shporer2009}; (9) \citet{Hartman2011a}; (10) \citet{Bakos2011}; (11) \citet{Ciceri2015}; (12) \citet{Quinn2010}; (13) \citet{Beky2011}; (14) \citet{Buchhave2011}; (15) \citet{Torres2012}; (16) \citet{Mortier2013}; (17) \citet{Bakos2012}; (18) \citet{Mancini2015}; (19) \citet{Sato2012}; (20) \citet{Hartman2012}; (21) \citet{Tsantaki2014}; (22) \citet{Boisse2013}; (23) \citet{Skillen2009}; (24) \citet{CollierCameron2010b}; (25) \citet{Faedi2011}; (26) \citet{Hellier2011}; (27) \citet{Bakos2015}; (28) \citet{Smith2012}; (29) \citet{Hebrard2013}; (30) \citet{West2016}; (31) \citet{Gillon2014}}
\tablenotetext{a}{The cited studies do not provide a $\log g$ measurement, so these numbers are computed from the quoted mass and radius values instead.}
\tablenotetext{b}{HAT-P-29 is part of the original FOHJ sample. This line is replicated from \citet{Ngo2015}.}
\end{deluxetable*}

\clearpage
\begin{deluxetable*}{llccc}
\tablecolumns{5}
\tablewidth{0pt}
\tablecaption{Flux ratio measurements of confirmed and candidate stellar companions  \label{tab:comp_phot}}
\tablehead{
\colhead{Companion\tablenotemark{a}} & \colhead{UT Obs. Date} & \colhead{$\Delta \Jfilt$} & \colhead{$\Delta \Hfilt$} & \colhead{$\Delta \Kfilt$} 
}
\startdata
HAT-P-5 cc  & 2013 Jul 04 & \nodata & \nodata & $6.71 \pm 0.15$ \\ 
HAT-P-5 cc  & 2015 Jun 24 & $6.84 \pm 0.21$ & \nodata & $6.669 \pm 0.073$ \\
HAT-P-27B & 2014 Jul 12 & $3.395 \pm 0.040$ & $3.107 \pm 0.021$ & $3.519 \pm 0.048$ \\
HAT-P-27B & 2015 Jan 09 & $3.3763 \pm 0.0093$ & $3.1436 \pm 0.0093$ & $3.520 \pm 0.011$ \\ 
HAT-P-27B & 2015 Jun 24 & \nodata & \nodata & $3.380 \pm 0.046$ \\
HAT-P-28 cc & 2015 Jul 07 & $3.333 \pm 0.025$ & \nodata & $3.168 \pm 0.040$ \\
HAT-P-28 cc\tablenotemark{b} & 2015 Jul 10 & $3.468 \pm 0.042$ & \nodata & $3.381 \pm 0.016$ \\ 
HAT-P-29B\tablenotemark{c} & 2012 Feb 02 & $7.09 \pm 0.15$ & \nodata & $6.92 \pm 0.16$ \\
HAT-P-29B & 2015 Jul 05 & \nodata & \nodata & $6.30 \pm 0.16$ \\
HAT-P-29B\tablenotemark{b} & 2015 Jul 10 & \nodata & \nodata & $6.85 \pm 0.18$ \\
HAT-P-35B & 2013 Mar 02 & $4.332 \pm 0.069$ & \nodata & $3.185 \pm 0.058$ \\ 
HAT-P-35B & 2014 Nov 10 & $3.726 \pm 0.025$ & $3.293 \pm 0.015$ & $3.562 \pm 0.032$ \\ 
HAT-P-39B & 2013 Mar 02 & $5.584 \pm 0.082$ & \nodata & $4.17 \pm 0.10$ \\ 
HAT-P-39B & 2014 Nov 07 & $4.686 \pm 0.050$ & $4.058 \pm 0.013$ & $4.40 \pm 0.16$ \\ 
HAT-P-41 cc & 2014 Oct 03 & \nodata & \nodata & $2.650 \pm 0.084$ \\ 
HAT-P-41 cc & 2015 Jun 24 & $2.947 \pm 0.017$ & \nodata & $2.527 \pm 0.045$ \\
HAT-P-54 cc & 2016 Jan 25 & \nodata & \nodata & $6.51 \pm 0.17$ \\
TrES-1 cc1\tablenotemark{d}   & 2013 Jul 04 & \nodata & \nodata & \nodata \\ 
TrES-1 cc1\tablenotemark{d}   & 2015 Jun 24 & \nodata & \nodata & \nodata \\ 
TrES-1 cc2   & 2013 Jul 04 & \nodata & \nodata & $6.676 \pm 0.060$ \\ 
TrES-1 cc2   & 2015 Jun 24 & $7.09 \pm 0.21$ & \nodata & $6.434 \pm 0.078$ \\
WASP-33 cc  & 2013 Aug 19 & $6.37 \pm 0.25$ & $5.71 \pm 0.12$ & $6.108 \pm 0.016$ \\ 
WASP-33 cc  & 2014 Dec 07 & \nodata & \nodata & $6.148 \pm 0.098$ \\
WASP-33 cc  & 2015 Dec 26 & \nodata & \nodata & $6.03 \pm 0.11$ \\
WASP-36 cc  & 2016 Jan 25 & \nodata & \nodata & $2.74 \pm 0.12$ \\ 
WASP-48 cc  & 2013 Aug 19 & \nodata & \nodata & $7.270 \pm 0.064$ \\
WASP-48 cc  & 2015 Jun 24 & $7.62 \pm 0.31$ & \nodata & $7.215 \pm 0.065$ \\ 
WASP-58B\tablenotemark{b}  & 2015 Jul 10 & $4.62 \pm 0.14$ & \nodata & $4.391 \pm 0.095$ \\
WASP-76B\tablenotemark{b}  & 2015 Jul 10 & $2.738 \pm 0.014$ & \nodata & $2.65 \pm 0.14$ \\
WASP-103 cc & 2016 Jan 25 & $2.427 \pm 0.030$ & $2.2165 \pm 0.0098$ & $1.965 \pm 0.019$ 
\enddata
\tablecomments{Except where noted, $\Delta\Kfilt$ is $\Delta\Ksfilt$.}
\tablenotetext{a}{We label companions with confirmed common proper motions as ``B'' and as ``cc'' when they are candidate companions. See Section~\ref{sec:companions}.}
\tablenotetext{b}{On 2015 Jul 10, we used the \Jcfilt\ and \BrGfilt\ bandpasses instead of \Jfilt\ and \Ksfilt, respectively. For these marked rows, $J$ corresponds to \Jcfilt\ and $K$ corresponds to \BrGfilt.}
\tablenotetext{c}{On 2012 Feb 02, for HAT-P-29, we used the \Kpfilt\ bandpass instead of \Ksfilt.}
\tablenotetext{d}{This candidate companion is too faint to obtain reliable photometric measurements.}
\end{deluxetable*}

\clearpage
\begin{deluxetable*}{llcccccc}
\tablecolumns{8}
\tabletypesize{\scriptsize}
\tablewidth{0pt}
\tablecaption{Multi-band photometry of confirmed and candidate stellar companions \label{tab:phot_colors}}
\tablehead{
\colhead{Companion\tablenotemark{a}} & \colhead{UT Obs. Date} & \colhead{$m_J$} & \colhead{$m_H$} & \colhead{$m_K$} & \colhead{$J-K$} & \colhead{$H-K$} & \colhead{$J-H$} 
}
\startdata
HAT-P-5 cc  & 2013 Jul 04 & \nodata & \nodata & $17.19 \pm 0.15$ & \nodata & \nodata & \nodata \\
HAT-P-5 cc  & 2015 Jun 24 & $17.68 \pm 0.21$ & \nodata & $17.150 \pm 0.073$ & $0.53 \pm 0.22$ & \nodata & \nodata \\
HAT-P-27B & 2014 Jul 12 & $14.021 \pm 0.040$ & $13.356 \pm 0.021$ & $13.628 \pm 0.048$ & $0.393 \pm 0.063$ & $-0.271 \pm 0.053$ & $0.664 \pm 0.045$ \\
HAT-P-27B & 2015 Jan 09 & $14.0023 \pm 0.0093$ & $13.3926 \pm 0.0093$ & $13.629 \pm 0.011$ & $0.374 \pm 0.014$ & $-0.236 \pm 0.014$ & $0.610 \pm 0.013$ \\
HAT-P-27B & 2015 Jun 24 & \nodata & \nodata & $13.489 \pm 0.046$ & \nodata & \nodata & \nodata \\
HAT-P-28 cc & 2015 Jul 07 & $14.894 \pm 0.025$ & \nodata & $14.272 \pm 0.040$ & $0.623 \pm 0.047$ & \nodata & \nodata \\
HAT-P-28 cc\tablenotemark{b} & 2015 Jul 10 & $15.029 \pm 0.042$ & \nodata & $14.485 \pm 0.016$ & $0.544 \pm 0.045$ & \nodata & \nodata \\
HAT-P-29B\tablenotemark{c} & 2012 Feb 02 & $17.74 \pm 0.15$ & \nodata & $17.22 \pm 0.16$ & $0.52 \pm 0.22$ & \nodata & \nodata \\
HAT-P-29B & 2015 Jul 05 & \nodata & \nodata & $16.60 \pm 0.16$ & \nodata & \nodata & \nodata \\
HAT-P-29B\tablenotemark{b} & 2015 Jul 10 & \nodata & \nodata & $17.15 \pm 0.18$ & \nodata & \nodata & \nodata \\
HAT-P-35B & 2013 Mar 02 & $15.690 \pm 0.069$ & \nodata & $14.215 \pm 0.058$ & $1.475 \pm 0.090$ & \nodata & \nodata \\
HAT-P-35B & 2014 Nov 10 & $15.084 \pm 0.025$ & $14.365 \pm 0.015$ & $14.592 \pm 0.032$ & $0.491 \pm 0.041$ & $-0.227 \pm 0.036$ & $0.718 \pm 0.029$ \\
HAT-P-39B & 2013 Mar 02 & $17.008 \pm 0.082$ & \nodata & $15.32 \pm 0.10$ & $1.68 \pm 0.13$ & \nodata & \nodata \\
HAT-P-39B & 2014 Nov 07 & $16.110 \pm 0.050$ & $15.242 \pm 0.013$ & $15.55 \pm 0.16$ & $0.56 \pm 0.17$ & $-0.31 \pm 0.16$ & $0.868 \pm 0.052$ \\
HAT-P-41 cc & 2014 Oct 03 & \nodata & \nodata & $12.378 \pm 0.084$ & \nodata & \nodata & \nodata \\
HAT-P-41 cc & 2015 Jun 24 & $12.953 \pm 0.017$ & \nodata & $12.255 \pm 0.045$ & $0.698 \pm 0.048$ & \nodata & \nodata \\
HAT-P-54 cc & 2016 Jan 25 & \nodata & \nodata & $16.84 \pm 0.17$ & \nodata & \nodata & \nodata \\
TrES-1 cc1\tablenotemark{d}   & 2013 Jul 04 & \nodata & \nodata & \nodata & \nodata & \nodata & \nodata\\
TrES-1 cc1\tablenotemark{d}   & 2015 Jun 24 & \nodata & \nodata & \nodata & \nodata & \nodata & \nodata \\
TrES-1 cc2  & 2013 Jul 04 & \nodata & \nodata & $16.495 \pm 0.060$ & \nodata & \nodata & \nodata \\
TrES-1 cc2   & 2015 Jun 24 & $17.38 \pm 0.21$ & \nodata & $16.253 \pm 0.078$ & $1.13 \pm 0.22$ & \nodata & \nodata \\
WASP-33 cc  & 2013 Aug 19 & $13.95 \pm 0.25$ & $13.22 \pm 0.12$ & $13.576 \pm 0.016$ & $0.38 \pm 0.25$ & $-0.35 \pm 0.12$ & $0.73 \pm 0.28$ \\
WASP-33 cc  & 2014 Dec 07 & \nodata & \nodata & $13.616 \pm 0.098$ & \nodata & \nodata & \nodata \\
WASP-33 cc  & 2015 Dec 26 & \nodata & \nodata & $13.49 \pm 0.11$ & \nodata & \nodata & \nodata \\
WASP-36 cc  & 2016 Jan 25 & \nodata & \nodata & $14.03 \pm 0.12$ & \nodata & \nodata & \nodata \\
WASP-48 cc  & 2013 Aug 19 & \nodata & \nodata & $17.642 \pm 0.064$ & \nodata & \nodata & \nodata \\
WASP-48 cc & 2015 Jun 24 & $18.25 \pm 0.31$ & \nodata & $17.587 \pm 0.065$ & $0.66 \pm 0.32$ & \nodata & \nodata \\
WASP-58B\tablenotemark{b}  & 2015 Jul 10 & $15.25 \pm 0.14$ & \nodata & $14.676 \pm 0.095$ & $0.57 \pm 0.17$ & \nodata & \nodata \\
WASP-76B\tablenotemark{b} & 2015 Jul 10 & $11.279 \pm 0.014$ & \nodata & $10.90 \pm 0.14$ & $0.38 \pm 0.14$ & \nodata & \nodata \\
WASP-103 cc & 2016 Jan 25 & $13.527 \pm 0.030$ & $13.0765 \pm 0.0098$ & $12.732 \pm 0.019$ & $0.795 \pm 0.035$ & $0.345 \pm 0.021$ & $0.450 \pm 0.031$ 
\enddata
\tablecomments{Except where noted, the \Kfilt\ bandpass used is the \Ksfilt\ bandpass. The $m_X$ columns report the secondary star's apparent magnitudes. The last three columns show the computed color of the companion star.}
\tablerefs{Primary star apparent magnitudes are from 2MASS~\citep{Skrutskie2006}.}
\tablenotetext{a}{We label companions with confirmed common proper motions as ``B'' and as ``cc'' when they are candidate companions. See Section~\ref{sec:companions}.}
\tablenotetext{b}{On 2015 Jul 10, we used the \Jcfilt\ and \BrGfilt\ bandpasses instead of \Jfilt\ and \Ksfilt, respectively. For these marked rows, $J$ corresponds to \Jcfilt\ and $K$ corresponds to \BrGfilt.}
\tablenotetext{c}{On 2012 Feb 02, for HAT-P-29, we used the \Kpfilt\ bandpass instead of \Ksfilt.}
\tablenotetext{d}{This candidate companion is too faint to obtain reliable photometric measurements.}
\end{deluxetable*}

\newpage
\begin{deluxetable*}{llcccc}
\tablecolumns{5}
\tablewidth{0pt}
\tablecaption{Astrometric measurements of all candidate stellar companions\label{tab:comp_astr}}
\tablehead{
& & & \colhead{$\rho$} & \colhead{PA} \vspace{-0.1cm}\\
\colhead{Candidate\tablenotemark{a}} & \colhead{UT Obs. Date} & \colhead{Band} &\vspace{-0.2cm}  & & \colhead{Reference}\\
& & & \colhead{(mas)} & \colhead{(\deg)}
}
\startdata
HAT-P-5 cc  & 2013 Jul 04 & \Ksfilt & $4313.7 \pm 2.7$ & $267.873 \pm 0.030$ & this work \\
HAT-P-5 cc  & 2015 Jun 24 & \Ksfilt & $4348.5 \pm 2.4$ & $267.557 \pm 0.032$ & this work\\
HAT-P-27B & 2013 Jun 27 & \ipfilt, \zpfilt & $ 656 \pm   21$ & $ 25.7 \pm   1.2$ & \citet{Woellert2015b}\\
HAT-P-27B & 2014 Jul 12 & \Ksfilt & $656.0 \pm 1.5$ & $25.48 \pm 0.13$ & this work \\
HAT-P-27B & 2015 Jan 09 & \Ksfilt & $653.9 \pm 1.5$ & $25.50 \pm 0.13$ & this work \\
HAT-P-27B & 2015 Mar 09 & \ipfilt, \zpfilt & $ 644 \pm    7$ & $ 28.4 \pm   1.9$ & \citet{Woellert2015b}\\
HAT-P-27B & 2015 Jun 24 & \Ksfilt & $652.8 \pm 1.5$ & $25.34 \pm 0.13$ & this work\\
HAT-P-28 cc & 2014 Oct 24 & \ipfilt, \zpfilt & $ 972 \pm   19$ & $212.3 \pm   2.0$ & \citet{Woellert2015b}\\
HAT-P-28 cc & 2015 Jul 07 & \Ksfilt & $996.6 \pm 1.5$ & $210.611 \pm 0.086$ & this work\\
HAT-P-28 cc & 2015 Jul 10 & \BrGfilt & $996.2 \pm 1.6$ & $210.614 \pm 0.088$ & this work\\
HAT-P-29B & 2012 Feb 02 & \Kpfilt & $3290.3 \pm 2.3$ & $159.892 \pm 0.032$ & this work \\
HAT-P-29B & 2014 Oct 21 & \ipfilt, \zpfilt & $3285 \pm   50$ & $161.5 \pm   2.4$ & \citet{Woellert2015b}\\
HAT-P-29B & 2015 Mar 06 & \ipfilt, \zpfilt & $3276 \pm  104$ & $160.7 \pm   1.4$ & \citet{Woellert2015b}\\
HAT-P-29B & 2015 Jul 05 & \Ksfilt & $3298.4 \pm 2.2$ & $159.558 \pm 0.033$ & this work\\
HAT-P-29B & 2015 Jul 10 & \BrGfilt & $3293.2 \pm 4.0$ & $159.572 \pm 0.040$ & this work\\
HAT-P-35B & 2013 Mar 02 & \Ksfilt & $932.1 \pm 1.5$ & $139.306 \pm 0.092$ & this work \\
HAT-P-35B & 2014 Apr 22 & $r_{TCI}$\tablenotemark{b} & $1016 \pm 11$ & $194.4 \pm 0.2$ & \citet{Evans2016}\\
HAT-P-35B & 2014 Nov 10 & \Ksfilt & $931.9 \pm 1.5$ & $139.270 \pm 0.090$ & this work \\
HAT-P-35B & 2015 Mar 09 & \ipfilt, \zpfilt & $ 933 \pm   10$ & $139.8 \pm   1.2$ & \citet{Woellert2015b}\\
HAT-P-39B & 2013 Mar 02 & \Ksfilt & $898.0 \pm 1.6$ & $94.31 \pm 0.10$ & this work \\
HAT-P-39B & 2014 Nov 07 & \Ksfilt & $900.4 \pm 1.7$ & $94.40 \pm 0.12$ & this work \\
HAT-P-41 cc & 2013 Jun 26 & \ipfilt, \zpfilt & $3619 \pm    5$ & $184.1 \pm   0.2$ & \citet{Woellert2015a}\\
HAT-P-41 cc & 2013 Apr 21 &  $r_{TCI}$\tablenotemark{b} & $3599 \pm 16$ & $183.7 \pm 0.2$ & \citet{Evans2016}\\
HAT-P-41 cc & 2014 Oct 03 & \Ksfilt & $3614.8 \pm 1.7$ & $184.102 \pm 0.026$ & this work \\
HAT-P-41 cc & 2014 Oct 21 & \ipfilt, \zpfilt & $3640 \pm   11$ & $184.0 \pm   0.1$ & \citet{Woellert2015b}\\
HAT-P-41 cc & 2015 Jun 24 & \Ksfilt & $3613.7 \pm 2.1$ & $184.094 \pm 0.031$ & this work\\
HAT-P-54 cc & 2014 Oct 21 & \ipfilt, \zpfilt & $4531 \pm 62$ & $135.95 \pm 1.96$ & \citet{Woellert2015b}\\
HAT-P-54 cc & 2015 Mar 06 & \ipfilt, \zpfilt & $4593 \pm 10$ & $135.82 \pm 0.27$ & \citet{Woellert2015b}\\ 
HAT-P-54 cc & 2016 Jan 25 & \Ksfilt & $4565.4 \pm 3.1$ & $135.652 \pm 0.035$ & this work\\
TrES-1 bg   & 2009 Jul 18-22\tablenotemark{c} & \ipfilt & $6190 \pm 30$ & $47.4 \pm 0.2$ & \citet{Faedi2013}\\
TrES-1 bg   & 2013 Jul 04 & \Ksfilt & $6355.2 \pm 2.1$ & $47.309 \pm 0.017$ & this work \\
TrES-1 bg   & 2015 Jun 24 & \Ksfilt & $6436.9 \pm 3.1$ & $47.321 \pm 0.024$ & this work\\
TrES-1 cc1  & 2013 Jul 04 & \Ksfilt & $2345.4 \pm 9.8$ & $172.91 \pm 0.11$ & this work \\
TrES-1 cc1  & 2015 Jun 24 & \Ksfilt & $2325.3 \pm 4.7$ & $171.71 \pm 0.078$ & this work \\ 
TrES-1 cc2 & 2009 Jul 18-22\tablenotemark{c} & \ipfilt & $4950 \pm 30$ & $149.6 \pm 0.5$ & \citet{Faedi2013}\\
TrES-1 cc2  & 2013 Jul 04 & \Ksfilt & $4940.2 \pm 2.2$ & $148.152 \pm 0.026$ & this work \\
TrES-1 cc2  & 2015 Jun 24 & \Ksfilt & $4946.5 \pm 2.6$ & $147.441 \pm 0.028$ & this work\\
WASP-33 cc & 2010 Nov 29 & \Kcfilt & $1961 \pm    3$ & $276.4 \pm   0.2$ & \citet{Moya2011}\\
WASP-33 cc  & 2013 Aug 19 & \Ksfilt & $1939.7 \pm 1.5$ & $276.247 \pm 0.045$ & this work \\
WASP-33 cc & 2014 Oct 21 & \ipfilt, \zpfilt & $1920 \pm   12$ & $275.9 \pm   0.7$ & \citet{Woellert2015b}\\
WASP-33 cc  & 2014 Dec 07 & \Ksfilt & $1934.3 \pm 1.6$ & $276.206 \pm 0.045$ & this work \\
WASP-33 cc  & 2015 Dec 26 & \Ksfilt & $1931.2 \pm 1.9$ & $276.350 \pm 0.058$ & this work\\
WASP-36 cc & 2014 Apr 23 &  $r_{TCI}$\tablenotemark{b} & $4872 \pm 19$ & $66.5 \pm 0.2$ & \citet{Evans2016}\\
WASP-36 cc & 2015 Mar 09 & \ipfilt, \zpfilt & $4845 \pm   17$ & $ 67.2 \pm   0.9$ & \citet{Woellert2015b}\\
WASP-36 cc  & 2016 Jan 25 & \Ksfilt & $4871.0 \pm 2.6$ & $66.921 \pm 0.028$ & this work\\
WASP-48 cc  & 2013 Aug 19 & \Ksfilt & $3571.9 \pm 2.6$ & $208.315 \pm 0.035$ & this work \\
WASP-48 cc  & 2015 Jun 24 & \Ksfilt & $3525.4 \pm 2.4$ & $209.053 \pm 0.037$ & this work\\
WASP-58B & 2013 Jun 25 & \ipfilt, \zpfilt & $1275 \pm   15$ & $183.2 \pm   0.4$ & \citet{Woellert2015a}\\
WASP-58B  & 2015 Jul 10 & \BrGfilt & $1286.0 \pm 1.6$ & $183.359 \pm 0.071$ & this work\\
WASP-76B & 2014 Aug 20 & \ipfilt & $443.8 \pm 5.3$ & $214.92 \pm 0.56$ & \citet{Ginski2016}\\
WASP-76B & 2014 Oct 21 & \ipfilt, \zpfilt & $ 425 \pm   12$ & $216.9 \pm   2.9$ & \citet{Woellert2015b}\\
WASP-76B  & 2015 Jul 10 & \BrGfilt & $442.5 \pm 1.5$ & $215.51 \pm 0.19$ & this work\\
WASP-103 cc & 2015 Mar 07 & \ipfilt, \zpfilt & $242 \pm 16 $ & $132.7 \pm 2.7$ & \citet{Woellert2015b}\\
WASP-103 cc & 2016 Jan 25 & \Ksfilt & $239.7 \pm 1.5$ & $131.41 \pm 0.35$ & this work
\enddata
\tablecomments{Separations ($\rho$) and position angle (PA) measurements of candidate companions in this work and other studies with published uncertainties. These values are plotted in Figures~\ref{fig:astr_plot1} and \ref{fig:astr_plot2}.}
\tablenotetext{a}{We label companions with confirmed common proper motions as ``B'', as ``cc'' when they are candidate companions, and as ``bg'' when they are confirmed background objects. See Section~\ref{sec:companions}.}
\tablenotetext{b}{The red filter used by \citet{Evans2016} is described as a combination of the Sloan \ipfilt\ and \zpfilt\ filters.}
\tablenotetext{c}{\citet{Faedi2013} did not provide a specific date for their observations. Here, we report the range of dates given and use the median value in our analysis.}
\end{deluxetable*}

\clearpage
\begin{deluxetable*}{llccccccc}
\tablecolumns{9}
\tabletypesize{\scriptsize}
\tablewidth{0pt}
\tablecaption{Derived stellar parameters of confirmed and candidate stellar companions\label{tab:interp_sec}}
\tablehead{
& &  \colhead{$T_{\mathrm{eff}}$} & \colhead{$M$} & \colhead{$\log g$} & \colhead{$D$} & \colhead{\Jfilt-band $T_{\mathrm{eff}}$} & \colhead{\Hfilt-band $T_{\mathrm{eff}}$} & \colhead{\Kfilt-band $T_{\mathrm{eff}}$} \vspace{-0.1cm}\\
\colhead{Companion\tablenotemark{a}} & \colhead{UT Obs. Date} & \vspace{-0.2cm} & & & & & & \\
& & \colhead{(K)} & \colhead{(\Msun)} & \colhead{(cgs)} & \colhead{(AU)} & \colhead{(K)} & \colhead{(K)} & \colhead{(K)}
}
\startdata
HAT-P-5 cc & 2013 Jul 04 & $2738 \pm 73$ & $0.0957 \pm 0.0043$ & $5.268 \pm 0.017$ & $718 \pm 62$ & \nodata & \nodata & $2738^{+70}_{-58}$ \\
HAT-P-5 cc & 2015 Jun 24 & $2814 \pm 72$ & $0.1009 \pm 0.0054$ & $5.248 \pm 0.019$ & $724 \pm 63$ & $2879^{+77}_{-70}$ & \nodata & $2754^{+32}_{-30}$ \\
HAT-P-27B & 2014 Jul 12 & $3460 \pm 45$ & $0.323 \pm 0.049$ & $4.941 \pm 0.039$ & $133.8 \pm 9.2$ & $3477.0^{+6.6}_{-5.9}$ & $3496.0^{+2.9}_{-3.7}$ & $3409.6^{+8.8}_{-7.3}$ \\
HAT-P-27B & 2015 Jan 09 & $3459 \pm 44$ & $0.323 \pm 0.048$ & $4.942 \pm 0.038$ & $133.4 \pm 9.2$ & $3479.9 \pm 1.5$ & $3490.2 \pm 1.5$ & $3409.6^{+2.2}_{-1.5}$ \\
HAT-P-27B & 2015 Jun 24 & $3433 \pm 23$ & $0.298 \pm 0.023$ & $4.968 \pm 0.019$ & $133.2 \pm 9.1$ & \nodata & \nodata & $3433.1^{+6.6}_{-7.3}$ \\
HAT-P-28 cc & 2015 Jul 07 & $3579 \pm 54$ & $0.444 \pm 0.043$ & $4.847 \pm 0.039$ & $394^{+34}_{-26}$ & $3609.3 \pm 5.7$ & \nodata & $3549.2 \pm 8.1$ \\
HAT-P-28 cc\tablenotemark{b} & 2015 Jul 10 & $3542 \pm 57$ & $0.409 \pm 0.050$ & $4.875 \pm 0.043$ & $394^{+34}_{-26}$ & $3583.3^{+9.8}_{-8.9}$ & \nodata & $3502.9 \pm 2.4$ \\
HAT-P-29B\tablenotemark{c} & 2012 Feb 02 & $2804 \pm 94$ & $0.1001 \pm 0.0069$ & $5.251 \pm 0.025$ & $1059^{+120}_{-69}$ & $2862^{+59}_{-50}$ & \nodata & $2749^{+72}_{-60}$ \\
HAT-P-29B & 2015 Jul 05 & $2955 \pm 78$ & $0.115 \pm 0.011$ & $5.206 \pm 0.028$ & $1062^{+120}_{-69}$ & \nodata & \nodata & $2955^{+54}_{-46}$ \\
HAT-P-29B\tablenotemark{b} & 2015 Jul 10 & $2710 \pm 110$ & $0.0942 \pm 0.0066$ & $5.274 \pm 0.027$ & $1060^{+120}_{-69}$ & \nodata & \nodata & $2713^{+89}_{-69}$ \\
HAT-P-35B & 2013 Mar 02 & $3525 \pm 76$ & $0.383 \pm 0.070$ & $4.889 \pm 0.059$ & $499 \pm 30$ & $3469.5^{+9.8}_{-8.9}$ & \nodata & $3583^{+13}_{-12}$ \\
HAT-P-35B & 2014 Nov 10 & $3563 \pm 70$ & $0.428 \pm 0.059$ & $4.859 \pm 0.051$ & $499 \pm 30$ & $3580.9 \pm 5.7$ & $3602.8 \pm 3.3$ & $3508.6^{+4.9}_{-5.7}$ \\
HAT-P-39B & 2013 Mar 02 & $3477 \pm 72$ & $0.324 \pm 0.068$ & $4.926 \pm 0.060$ & $576^{+100}_{-59}$ & $3413 \pm 13$ & \nodata & $3548^{+21}_{-20}$ \\
HAT-P-39B & 2014 Nov 07 & $3558 \pm 52$ & $0.422 \pm 0.044$ & $4.862 \pm 0.038$ & $577^{+100}_{-59}$ & $3558 \pm 11$ & $3614.2^{+2.4}_{-3.3}$ & $3504^{+32}_{-24}$ \\
HAT-P-41 cc & 2014 Oct 03 & $3783 \pm 67$ & $0.561 \pm 0.028$ & $4.737 \pm 0.026$ & $1124^{+130}_{-98}$ & \nodata & \nodata & $3783^{+35}_{-29}$ \\
HAT-P-41 cc & 2015 Jun 24 & $3873 \pm 83$ & $0.593 \pm 0.028$ & $4.707 \pm 0.024$ & $1124^{+130}_{-98}$ & $3914.1^{+8.9}_{-8.1}$ & \nodata & $3834^{+21}_{-20}$ \\
HAT-P-54 cc\tablenotemark{d} & 2016 Jan 25 & $1941 \pm 75$ & $0.07428 \pm 0.00090$ & $5.3996 \pm 0.0083$ & $619 \pm 16$ & \nodata & \nodata & $1941^{+78}_{-62}$ \\
TrES-1 cc1\tablenotemark{e} & 2013 Jul 04 & \nodata  & \nodata  & \nodata  & \nodata & \nodata & \nodata &\nodata  \\
TrES-1 cc1\tablenotemark{e} & 2015 Jun 24 &  \nodata  & \nodata  & \nodata  & \nodata & \nodata & \nodata &\nodata  \\
TrES-1 cc2 & 2013 Jul 04 & $2550 \pm 140$ & $0.0874 \pm 0.0047$ & $5.307 \pm 0.025$ & $641 \pm 43$ & \nodata & \nodata & $2554^{+27}_{-26}$ \\
TrES-1 cc2 & 2015 Jun 24 & $2580 \pm 170$ & $0.0884 \pm 0.0061$ & $5.301 \pm 0.031$ & $642 \pm 43$ & $2507^{+110}_{-86}$ & \nodata & $2661^{+32}_{-30}$ \\
WASP-33 cc & 2013 Aug 19 & $3256 \pm 59$ & $0.183 \pm 0.024$ & $5.081 \pm 0.030$ & $239 \pm 14$ & $3276^{+52}_{-47}$ & $3316^{+22}_{-20}$ & $3181.0 \pm 4.1$ \\
WASP-33 cc & 2014 Dec 07 & $3171 \pm 29$ & $0.1560 \pm 0.0084$ & $5.124 \pm 0.012$ & $238 \pm 14$ & \nodata & \nodata & $3171^{+25}_{-24}$ \\
WASP-33 cc & 2015 Dec 26 & $3201 \pm 31$ & $0.1650 \pm 0.0093$ & $5.111 \pm 0.013$ & $238 \pm 14$ & \nodata & \nodata & $3201^{+29}_{-25}$ \\
WASP-36 cc & 2016 Jan 25 & $3583 \pm 67$ & $0.451 \pm 0.053$ & $4.846 \pm 0.048$ & $2190 \pm 580$ & \nodata & \nodata & $3583^{+28}_{-24}$ \\
WASP-48 cc & 2013 Aug 19 & $2768 \pm 51$ & $0.0974 \pm 0.0030$ & $5.261 \pm 0.012$ & $1660 \pm 180$ & \nodata & \nodata & $2768^{+28}_{-26}$ \\
WASP-48 cc & 2015 Jun 24 & $2810 \pm 50$ & $0.1001 \pm 0.0036$ & $5.251 \pm 0.013$ & $1640 \pm 170$ & $2830^{+120}_{-110}$ & \nodata & $2792^{+28}_{-27}$ \\
WASP-58B\tablenotemark{b} & 2015 Jul 10 & $3396 \pm 53$ & $0.265 \pm 0.042$ & $4.997 \pm 0.040$ & $384 \pm 64$ & $3419^{+23}_{-22}$ & \nodata & $3374^{+16}_{-18}$ \\
WASP-76B\tablenotemark{b} & 2015 Jul 10 & $4310 \pm 170$ & $0.712 \pm 0.042$ & $4.608 \pm 0.030$ & $53.0 \pm 8.8$ & $4486.3^{+9.8}_{-8.1}$ & \nodata & $4155^{+98}_{-80}$ \\
WASP-103 cc & 2016 Jan 25 & $4330 \pm 100$ & $0.721 \pm 0.024$ & $4.604 \pm 0.016$ & $112.4 \pm 8.4$ & $4369 \pm 21$ & $4252.2 \pm 6.5$ & $4374 \pm 16$ 
\enddata
\tablecomments{For each date, we report error weighted averages of all measurements on $T_{\mathrm{eff}}$, $M$, $\log g$, and $D$. Our uncertainties account for uncertainties arising from the measurements, the primary star's stellar parameters and the error weighted average calculation.  However, they do not include uncertainties from the stellar models and our assumptions on stellar composition. All uncertainties are thus underestimates of the true uncertainty, especially for the final three columns, as these only include measurement uncertainties. The last three columns also show that if our candidate companions are comoving, their temperatures are consistent with a late type main sequence star in all filters. Except where noted, \Kfilt\ corresponds to \Ksfilt.}
\tablenotetext{a}{We label companions with confirmed common proper motions as ``B'' and as ``cc'' when they are candidate companions. See Section~\ref{sec:companions}.}
\tablenotetext{b}{On 2015 Jul 10, we used the \Jcfilt\ and \BrGfilt\ bandpasses instead of \Jfilt\ and \Ksfilt, respectively. Therefore, for these marked rows, the seventh and ninth columns report the effective temperature for these bandpasses instead.}
\tablenotetext{c}{On 2012 Feb 02, for HAT-P-29, we used the \Kpfilt\ bandpass instead of \Ksfilt.}
\tablenotetext{d}{For this target, the companion temperature is below the lower limit of the PHOENIX models (2300\unit{K}), so we assumed a blackbody for both primary and secondary stars.}
\tablenotetext{e}{This candidate companion is too faint to obtain reliable photometric measurements.}
\end{deluxetable*}

%

\end{document}